\documentclass[ejs, preprint]{imsart}
\usepackage{amsmath, amsthm}
\usepackage{mathtools}
\usepackage{tikz}
\usepackage{amsfonts}
\usepackage{bbm}
\usepackage{natbib}
\usepackage[colorlinks,citecolor=blue,urlcolor=blue,filecolor=blue,backref=page]{hyperref}
\usepackage{graphicx}
\usepackage{float}
\usepackage{booktabs}
\usepackage{threeparttable}
\usepackage{listings}

\usepackage{booktabs}  
\usepackage{multirow}  

\startlocaldefs
\graphicspath{ {./images/} }

\newcommand{\D}{\text{D}}
\newcommand{\I}{\text{I}}

\newcommand{\mbe}{\mathbb{E}}

\newcommand{\KL}[2]{\text{KL}\left(  #1 \, || \, #2 \right)}

\newcommand{\normal}{\mathcal{N}}

\newcommand{\rmse}{\text{RMSE}}

\newcommand{\elpd}{\text{ELPD}}

\newtheorem{proposition}{Proposition}

\newcommand{\myosfresults}{\url{https://osf.io/fuean/}}

\usetikzlibrary{shapes,fit} 
\usetikzlibrary{bayesnet}
\usetikzlibrary{shapes, shapes.geometric, decorations,arrows,calc,arrows.meta,fit,positioning}
\tikzset{
    -Latex,auto,node distance =1 cm and 1 cm,semithick,
    state/.style ={circle, draw, minimum width = 1.0 cm},
    point/.style = {circle, draw, inner sep=0.04cm,fill,node contents={}}, 
    vmissing/.style={
    draw=none, 
    scale=1,
    text height=0.111cm,
    execute at begin node=\color{black}$\vdots$
    },
    hmissing/.style={
    draw=none, 
    scale=1,
    text height=0.111cm,
    execute at begin node=\color{black}$...$
    },
    bidirected/.style={Latex-Latex,dashed},
    el/.style = {inner sep=2pt, align=left, sloped}
}

\tikzset{
  startstop/.style={rectangle, rounded corners, minimum width=3cm, minimum height=1cm,text centered, draw=black, fill=gray!20},
  process/.style={rectangle, minimum width=3.5cm, minimum height=1cm, text centered, draw=black, fill=blue!10},
  decision/.style={diamond, aspect=2, minimum width=3.5cm, minimum height=1cm, text centered, draw=black, fill=orange!20},
  arrow/.style={thick, -{Stealth}}
}

\endlocaldefs

\newcommand*{\getFirstNameFirstAuthor}{Javier Enrique}
\newcommand*{\getLastNameFirstAuthor}{Aguilar}

\newcommand*{\getFirstNameSecondAuthor}{Paul-Christian}
\newcommand*{\getLastNameSecondAuthor}{Bürkner}

\newcommand*{\getRunAuthor}{J.E. Aguilar and P-C. Bürkner }

\newcommand*{\getTitle}{Dependency-Aware Shrinkage Priors for High Dimensional Regression}
\newcommand*{\getRunningTitle}{Dependency-Aware Shrinkage Priors}

\newcommand*{\getFirstAddress}{Department of Statistics, TU Dortmund University, Germany}

\newcommand*{\getMailFirstAuthor}{javier.aguilarr@icloud.com}
\newcommand*{\getMailSecondAuthor}{paul.buerkner@gmail.com}

\newcommand*{\getURLFirstAuthor}{https://jear2412.github.io}
\newcommand*{\getURLSecondAuthor}{https://paul-buerkner.github.io}

\pubyear{2025}
\volume{TBA}
\issue{TBA}
\firstpage{1}
\lastpage{1}

\begin{document}

\begin{frontmatter}

\title{ \getTitle  }
\runtitle{ \getRunningTitle}
\date{28.01.25}

\begin{aug}
\author{\fnms{\getFirstNameFirstAuthor} \snm{\getLastNameFirstAuthor}\thanksref{addr1,t1}\ead[label=e1]{\getMailFirstAuthor}%
\ead[label=u1,url]{\getURLFirstAuthor}
}
\and
\author{\fnms{\getFirstNameSecondAuthor} \snm{\getLastNameSecondAuthor}\thanksref{addr1,t2}\ead[label=e2]{\getMailSecondAuthor}%
\ead[label=u2,url]{\getURLSecondAuthor}}

\runauthor{ \getRunAuthor}
\address[addr1]{\getFirstAddress}

\thankstext{t1}{Corresponding author: \printead{e1} \printead{u1}}

\thankstext{t2}{\printead{e2} \printead{u2}}
\end{aug}

\begin{abstract}
\noindent
In high dimensional regression, global local shrinkage priors have gained significant traction for their ability to yield sparse estimates, improve parameter recovery, and support accurate predictive modeling. While recent work has explored increasingly flexible shrinkage prior structures, the role of explicitly modeling dependencies among coefficients remains largely unexplored. In this paper, we investigate whether incorporating such structures into traditional shrinkage priors improves their performance. We introduce dependency-aware shrinkage priors, an extension of continuous shrinkage priors that integrates correlation structures inspired by Zellner’s $g$ prior approach. We provide theoretical insights into how dependence alters the prior and posterior structure, and evaluate the method empirically through simulations and real data. We find that modeling dependence can improve parameter recovery when predictors are strongly correlated, but offers only modest gains in predictive accuracy. These findings suggest that prior dependence should be used selectively and guided by the specific inferential goals of the analysis.

\end{abstract}


\end{frontmatter}


\section{Introduction}

Regression analysis has long been a cornerstone of statistical modeling \citep{gelman_data_2006,hastie_elements_2009,hastie_statistical_2015, gelman_regression_2020}. As datasets have grown in size and complexity, the need for models capable of handling high dimensional settings while avoiding overfitting has become increasingly evident \citep{buehlmann_high-dimensional_2014, giraud_introduction_2014, vershynin_high-dimensional_2018}. To address this challenge, a wide array of priors with strong theoretical and empirical properties has been proposed over the past decade \citep{carvalho_horseshoe_2010,griffin_inference_2010,bhattacharya_dirichletlaplace_2015,piironen_sparsity_2017,rockova_spike-and-slab_2018,zhang_bayesian_2020}.

Let $y_i \in \mathbb{R}$ denote the $i$th response variable, related to $p$ explanatory variables $x_i' = (x_{i1}, \ldots, x_{ip}) \in \mathbb{R}^p$ through the linear regression model
\begin{equation}
\label{eq:linregeq}
y_i = \alpha + x_i' b + \varepsilon_i, \quad i = 1, \ldots, n,    
\end{equation}
where $\alpha \in \mathbb{R}$ is the intercept, $b = (b_1, \ldots, b_p)' \in \mathbb{R}^p$ is the vector of regression coefficients, and $\varepsilon_i$ is the error term, assumed to satisfy $\varepsilon_i \sim \mathcal{N}(0, \sigma^2)$ with unknown residual variance $\sigma^2$ \citep{gelman_bayesian_2013}.

Although model \eqref{eq:linregeq} is standard, we are particularly interested in the high dimensional regime where $p > n$ \citep{tibshirani_regression_1996,giraud_introduction_2014,wainwright_high-dimensional_2019}. In this setting, a central challenge is the specification of prior distributions for the coefficients $b$ and the residual variance $\sigma^2$ \citep{gelman_prior_2006, tadesse_handbook_2021}. Introducing additional assumptions about the data-generating process can simplify this task. In particular, if we assume that the true model is sparse—that is, only a small subset of the coefficients are nonzero—then shrinkage priors become an attractive modeling choice \citep{johnstone_needles_2004,castillo_needles_2012,pas_conditions_2016, pas_uncertainty_2017}. Sparse solutions are heavily recommended: dense models not only hinder interpretability and statistical parsimony, but may also lead to computational inefficiencies and overfitting \citep{buhlmann_statistics_2011, giraud_introduction_2014, hastie_statistical_2015, wainwright_high-dimensional_2019}.

Among shrinkage priors, the class of continuous global-local shrinkage priors has gained significant attention \citep{mitchell_bayesian_1988,carvalho_horseshoe_2010, armagan_generalized_2013,bhattacharya_dirichletlaplace_2015,pas_conditions_2016, bhadra_default_2016, zhang_bayesian_2020, tadesse_handbook_2021}. These priors balance sparsity and flexibility by combining global shrinkage, which controls the overall level of shrinkage, with local shrinkage, which allows individual coefficients to escape strong penalization. Their popularity stems from a combination of excellent theoretical properties—such as minimax optimality and desirable posterior concentration rates—and strong empirical performance \citep{castillo_needles_2012, castillo_bayesian_2015,pas_conditions_2016, ghosh_asymptotic_2017, van_der_pas_theoretical_2021, tadesse_handbook_2021}. Furthermore, their compatibility with efficient computational strategies and probabilistic programming frameworks has contributed to their widespread adoption \citep{george_variable_1993, piironen_sparsity_2017,bhattacharya_fast_2016,aguilar_intuitive_2023}.

Incorporating prior knowledge about regression coefficients can, in principle, improve both parameter estimation and predictive performance, provided that the prior sufficiently reflects true underlying structure \citep{pas_uncertainty_2017, simpson_penalising_2017}. For instance, modeling multivariate dependencies among coefficients allows to capture joint uncertainty in related parameters \citep{gelman_bayesian_2013, gelman_regression_2020}. While such approaches have been explored in non-shrinkage contexts, analogous developments within shrinkage prior frameworks remain relatively limited, since conditional independence of the coefficients is typically assumed \citep{zellner_models_1996, agliari_g_1988, casella_penalized_2010, griffin_priors_2013, griffin_structured_2024}.

A foundational example of priors that incorporate dependencies is Zellner’s $g$ prior, which uses the design matrix to encode the dependence structure of the regression coefficients via their prior covariance matrix \citep{zellner1986assessing, maruyama_fully_2011, li_mixtures_2018}. Although not originally conceived as a shrinkage prior, the $g$ prior exhibits a global shrinkage structure and remains widely used in Bayesian model selection and averaging due to its closed-form marginal likelihood \citep{robert_monte_2004,liang_mixtures_2008,maruyama_fully_2011,li_mixtures_2018}.

Some approaches attempt to encode coefficient dependence by treating regularized objectives as priors; specifically, by normalizing expressions of the form $b'\Omega b + R(b)$, where $\Omega$ is a positive semidefinite matrix and $R(b)$ is a penalty term \citep{casella_penalized_2010,pauger_bayesian_2019,goplerud_modelling_2021}. Choosing $R(b)$ as a convex penalty leads to computationally tractable optimization problems \citep{tibshirani_regression_1996, hastie_statistical_2015}. However, translating these penalties into fully Bayesian priors is not straightforward: the role of hyperparameters and their interaction with the likelihood can be difficult to interpret or tune \citep{casella_penalized_2010, hahn_decoupling_2015, griffin_structured_2024}. Moreover, the key advantages of regularization methods, namely computational efficiency and the ability to produce sparse solutions, are often diminished in the Bayesian setting \citep{hahn_decoupling_2015}.

Additionally, although these constructions superficially resemble frequentist regularization techniques (such as interpreting the log-prior as a penalty), the comparison can be misleading. The theoretical properties of Bayesian shrinkage priors and frequentist regularizers differ in fundamental ways \citep{simpson_penalising_2017, castillo_bayesian_2024}. For example, under suitable conditions on the design matrix, the Lasso estimator can consistently recover sparse signals \citep{zhao_model_2006}. In contrast, the Laplace prior, despite its formal similarity to the Lasso penalty, tends to over-shrink coefficients due to its insufficient tail mass, often leading to biased estimates and poor uncertainty quantification \citep{castillo_bayesian_2015,pas_uncertainty_2017,bhadra_lasso_2019, castillo_bayesian_2024}. While initial analogies between Bayesian and frequentist approaches may be instructive, they can obscure important differences in parameter recovery, shrinkage behavior, and predictive performance.

A more recent and notable development is the Structured Shrinkage Prior proposed by \citet{griffin_structured_2024}, which incorporates dependence by constructing a covariance matrix as the elementwise product of the second-moment matrix of the local scales and an arbitrary structure-imposing matrix. While this framework generalizes existing approaches and offers conceptual elegance, it is strongly limited to shrinkage priors with unit second moments for the local scales. This constraint is nontrivial, as the second moment is sensitive to the choice of hyperparameters—an important tuning mechanism in high dimensional models—and may not equal one in practice. See Section \ref{sec:methods} for further details.

While prior dependence structures have been explored in low dimensional asymptotics, their role in high dimensional models remains poorly understood. For instance, \citet{hagar_posterior_2024} show that in low dimensions, complex dependence encoded through copula-based priors may be overridden by the likelihood as sample size increases. However, their results rely on the Bernstein–von Mises theorem and do not apply in high dimensional settings, where the behavior of posterior dependence structures remains largely an open question \citep{ghosal_convergence_2000,ghosh_bayesian_2003,castillo_bayesian_2015}.

\subsection{Contributions of this paper}

The main aim of this paper is to investigate whether traditional continuous global-local shrinkage priors can benefit from the inclusion of dependency structures in high dimensional sparse settings, and to determine the conditions under which such benefits arise. To this end, we introduce a generalization of global-local shrinkage priors, which we term dependency-aware shrinkage priors (DASP). Our approach extracts correlation information from the design matrix of the predictors and combines it with the shrinkage scales in a data driven manner. 

To evaluate the impact of incorporating dependence structures, we present both theoretical and empirical results. We characterize how introducing dependence modifies the structure of the prior and posterior distributions. We conduct simulation studies using both the true correlation matrix and our data-driven approach. The results show that incorporating dependence can improve parameter recovery when strong correlation exists among groups of coefficients. However, the gains in predictive performance are generally modest. We further validate our method on multiple real-world datasets, where we observe consistent patterns: dependency-aware priors may aid estimation in structured settings but do not yield substantial improvements in prediction.

We conclude that, while prior dependence structures can be useful in specific inferential contexts, they are not universally beneficial and should not be used by default. In high dimensional problems, the advantages of flexible prior dependence appear modest and highly context dependent. Our conclusion is that such structures should be employed judiciously, with clear justification based on the goals of the analysis.

\section{Methods}
\label{sec:methods}

\subsection{Shrinkage priors}
\label{subsec:shrinkage_priors}
Continuous global-local shrinkage priors are constructed as scale mixtures of normal distributions \citep{west_scale_1987}, yielding a hierarchical model for the regression coefficients. Each coefficient $b_i$ is modeled as
\begin{equation}
\label{eq:spprior}
b_i \mid \sigma, \tau, \lambda_i \sim \normal(0, \sigma^2 \tau^2 \lambda_i^2), \quad \lambda_i \sim p(\lambda_i), \quad \tau \sim p(\tau), \quad \sigma \sim p(\sigma),
\end{equation}
where $\tau$ is a global scale parameter that controls the overall level of shrinkage, and $\lambda_i$ are local scale parameters that allow coefficient-specific deviations. The global scale reflects the belief that most coefficients are near zero, while the local scales allow signals to escape shrinkage when warranted by the data. This hierarchy encourages sharing of information about sparsity across coefficients while retaining flexibility at the individual level. We include the residual variance $\sigma^2$ within the prior variance of $b_i$, following standard practice, as this improves adaptivity to varying signal-to-noise ratios \citep{van_der_pas_theoretical_2021}.

With appropriate choices for the priors on $\lambda_i$ and $\tau$, global-local shrinkage priors induce approximate sparsity, shrinking irrelevant signals strongly toward zero while avoiding excessive shrinkage of relevant signals. The combination of local and global scales also endows these priors with desirable theoretical properties, such as minimax optimality, posterior consistency, and optimal contraction rates under sparsity \citep{castillo_needles_2012,ghosh_asymptotic_2017,pas_conditions_2016, van_der_pas_theoretical_2021}. Although they do not produce exact zeros, which may be required in some applications, model selection can be handled post hoc by decoupling inference from selection  \citep{hahn_decoupling_2015, piironen_projective_2020}.

The specific choices for the priors on $\lambda_i$ and $\tau$ give rise to the wide array of shrinkage priors used today. Common examples include the Horseshoe, Three-Parameter Beta, Normal-Gamma, and Beta Prime priors, among others \citep{carvalho_horseshoe_2010, armagan_generalized_2011, griffin_inference_2010, bai_large-scale_2019}. These models belong to the family of shrinkage priors that treat the local scales $\lambda_i$ as conditionally independent. More recently, alternative approaches have been proposed that introduce joint modeling of the local scales via a multivariate distribution; examples of this include the Dirichlet–Laplace, R2D2, and Generalized R2 Decomposition priors \citep{bhattacharya_dirichletlaplace_2015, zhang_bayesian_2020, aguilar_generalized_2025}.

\subsection{Dependency-aware shrinkage priors}
\label{subsec:dasp}

Regression coefficients are typically deemed as conditionally independent a priori under the shrinkage prior setup \eqref{eq:spprior}. We propose a natural extension of this model by considering dependencies in the regression coefficients via correlation matrix $\Omega$. The model we propose is following:
\begin{equation}
\label{eq:dasp-prior}
     b \, \mid  \, \sigma, \tau, \lambda \sim \normal \left( 0, \sigma^2  \tau^2  \D_\lambda \, \Omega \, \D_\lambda  \right), \ \ \lambda_i \sim p(\lambda_i), \ \ \tau \sim p(\tau), \ \ \sigma \sim p(\sigma),
\end{equation}
where $\D_\lambda$ is the diagonal matrix that contains the scales $\lambda_i$ as diagonal elements and $\Omega$ is a correlation matrix. We refer to priors of this form \eqref{eq:dasp-prior} as dependency-aware shrinkage priors (DASP). The standard shrinkage prior model  \eqref{eq:spprior} is recovered by setting $\Omega = \I$. 

\subsubsection{Related priors}
\label{sec:other_priors}

Priors for regression coefficients that capture dependence structures typically do so through the incorporation of a covariance matrix. In what follows, we discuss two such priors: Zellner’s $g$ prior and Structured Shrinkage Priors (SSPs), emphasizing their respective advantages and identifying their limitations. We argue that the prior introduced in \eqref{eq:dasp-prior} offers a natural and robust extension that directly addresses these limitations.

Zellner’s $g$ prior \citep{zellner1986assessing, agliari_g_1988, maruyama_fully_2011, li_mixtures_2018} incorporates dependence information directly from the design matrix $X$ by specifying
\begin{equation}
\label{eq:zellner-prior}
b \mid \sigma, g  \sim \normal \left( 0, \sigma^2 g (X'X)^{-1} \right), \quad
\sigma \sim p(\sigma),
\end{equation}
where $g > 0$ is either fixed or endowed with a prior distribution, and $X$ is the design matrix composed of the covariate vectors $x_i'$. This formulation ensures that the prior covariance of $b$ is proportional to the covariance of the maximum likelihood estimator (MLE) $\hat{b}$, thus preserving scale invariance with respect to the regressors \citep{hoff_first_2009}. Zellner’s prior is widely appreciated for its analytical tractability, particularly in Bayesian model selection. When $g$ is fixed, it yields closed-form expressions for the marginal likelihood and posterior, thereby eliminating the need for sampling-based approximations \citep{robert_monte_2004, maruyama_fully_2011, li_mixtures_2018}. 

Within this framework, $g$ functions as a global shrinkage parameter, applying uniform regularization across all components of $b$. Although $(X'X)^{-1}$ provides some localized scaling, the $g$ prior lacks adaptivity, as it does not include coefficient-specific local scales $\lambda_i$ \citep{liang_mixtures_2008, gelman_bayesian_2013}. As a result, fixed or poorly chosen values of $g$ can lead to over-shrinkage of relevant variables or under-penalization of noise. While various priors for $g$ have been proposed to mitigate this \citep{liang_mixtures_2008, maruyama_fully_2011, li_mixtures_2018}, the absence of local scales ultimately excludes the $g$ prior from the class of adaptive shrinkage priors.

In high dimensional settings, the $g$ prior is further limited by its reliance on $(X'X)^{-1}$, making it sensitive to collinearity and rank deficiency. Common remedies include replacing $X'X$ with $(X'X + \eta I)$ for some $\eta > 0$, though this introduces an arbitrary tuning parameter. Alternatively, the Moore–Penrose pseudoinverse can be used, but this yields an improper prior with support limited to the column space of $X$, leading to an improper posterior. Reparametrization via QR or SVD decomposition offers a more a numerically stable alternative by restricting the prior to the column space of $X$, but this comes at the cost of interpretability with respect to the original covariates.

By contrast, our formulation \eqref{eq:dasp-prior} is capable of mimicking the $g$ prior through appropriate specification of $\Omega$, as shown in Section~\ref{subsec:dep_structs}. Moreover, the inclusion of local scale parameters $\lambda_i$ directly addresses the issue of adaptivity \citep{liang_mixtures_2008}, offering a flexible and robust alternative in both low and high dimensional settings.

In the broader context of shrinkage priors, a structurally related idea appears in the work of \citet{george_variable_1993, george_stochastic_1995, george_approaches_1997}, who propose the spike-and-slab prior with a non-diagonal precision matrix to encode dependencies among coefficients. In their formulation, the prior on $b$ takes the form $\mathcal{N}(0, \tau^2 \Gamma \Omega^{-1} \Gamma)$, where $\Gamma$ is a diagonal matrix of binary inclusion indicators $\gamma_i \in \{0,1\}$ for $i = 1,\ldots,p$. Their work suggests using $\Omega = (X'X)^{-1}_\gamma$, where $X_\gamma$ is the submatrix of $X$ corresponding to the active variables, and also explores special cases such as $\Omega = I$ and $\Omega = (X'X)^{-1}$, independent of $\gamma$. However, in practice, their implementations focus on mixtures of normals with diagonal $\Omega$, retaining conditional independence among coefficients which gives rise to the typical spike-and-slab prior. Our formulation differs in several key respects: we consider continuous local scales $\lambda_i$ instead of binary indicators, yielding a fully continuous global-local prior that avoids discrete model search \citep{hahn_decoupling_2015, tadesse_handbook_2021}, and we construct $\Omega$ from the full design matrix or a structured estimate thereof, allowing scalable and adaptive shrinkage in high dimensional settings. We show the latter in Section \ref{subsec:dep_structs}.

Turning to continuous global-local formulations, \citet{griffin_structured_2024} propose the Structured Shrinkage Priors (SSPs) framework, which introduces dependence structures via a product representation of normal scale mixtures. Specifically, if $z \sim \normal(0, \Phi)$ and $\lambda$ is a vector of stochastic scales independent of $z$, then the elementwise product $b = \lambda \circ z$ defines a scale mixture of normals \citep{west_scale_1987}, with prior covariance:
\begin{equation}
\label{eq:sp_cov}
\Sigma_b = \mathbb{E}(\lambda \lambda') \circ \Phi.
\end{equation}
In standard settings, $\Phi = I$ yields unstructured shrinkage priors. However, when $\Phi \ne I$, identifiability issues arise: the individual scales $\lambda_i$ are not separately identifiable from the diagonal entries of $\Phi$. To address this, \citet{griffin_structured_2024} impose the constraint $\mathbb{E}(\lambda_i^2) = 1, i = 1, \dots, p$. While this restores identifiability, it significantly restricts the flexibility of the prior by limiting the range of hyperparameter choices available for the distribution of $\lambda_i$.

For instance, under this constraint, the Normal-Gamma prior \citep{griffin_inference_2010}, with $\lambda^2 \sim \text{Gamma}(\alpha, \beta)$, is only admissible when $\alpha = \beta$. Similarly, heavy-tailed priors such as the Beta Prime \citep{johnson_continuous_1994}, which are often used to induce marginal heavy tails in $b$, become infeasible unless their parameter values satisfy the constraint \citep{pas_conditions_2016, ghosh_asymptotic_2017, bai_large-scale_2019, zhang_bayesian_2020, aguilar_intuitive_2023}. This constraint also prevents straightforward generalizations of well-established priors like the Horseshoe \citep{carvalho_horseshoe_2010}, since it assumes $\lambda_i \sim \mathcal{C}^+(0, 1)$, implying that $\mathbb{E}(\lambda_i^2)$ does not exist.

Finally, we observe that the matrix $\Phi$ in the formulation of Structured Shrinkage Priors (SSPs) remains relatively underexplored, even though it plays a critical role in determining the properties of the model. As \citet{griffin_structured_2024} illustrate, altering $\Phi$ within traditional shrinkage priors can lead to a notable increase in computational cost, emphasizing the importance of a more systematic study of its effects. In addition, the task of specifying $\Phi$ is typically left to the user, either through direct selection or by placing a prior distribution on it; both of which present open opportunities for further investigation.

In contrast, our dependency-aware shrinkage prior avoids imposing any constraint on the second moment of the local scales $\lambda_i$. This flexibility allows it to encompass a broad class of shrinkage priors, including those with undefined or heavy-tailed moments, as long as $\Omega$ is appropriately specified. In the following sections, we analyze some properties of our proposed prior and examine the practical consequences of incorporating $\Omega$.

\subsubsection{Conditional means}
\label{subsec:cond_means}

First, we discuss how the conditional posterior mean is affected by the presence of $\Omega$. The conditional posterior distribution of $b$ under our dependency-aware shrinkage prior \eqref{eq:dasp-prior} is multivariate Gaussian. Its mean and covariance matrix are given by
\begin{equation}
\label{eq:pc-var}
\mathbb{E}(b \mid y, \lambda, \tau, \sigma, \Omega) = Q_\Omega^{-1} X' y, \
\mathrm{Cov}(b \mid y, \lambda, \tau, \sigma, \Omega) = \sigma^2 Q_\Omega^{-1},
\end{equation}
where $Q_\Omega \coloneqq X'X + 1/\tau^2 D_\lambda^{-1} \Omega^{-1} D_\lambda^{-1}.$
We explicitly condition on $\Omega$ to highlight its role in shaping both the posterior mean and covariance structure. Assuming that $X$ is of full rank, the posterior mean can be alternatively expressed in terms of the MLE $\hat{b}$ as
\begin{equation}
\label{eq:pc-mean-mle}
\mathbb{E}(b \mid y, \lambda, \tau, \sigma, \Omega) = \tau^2 D_\lambda \Omega D_\lambda \left( \tau^2 D_\lambda \Omega D_\lambda + (X'X)^{-1} \right)^{-1} \hat{b}.
\end{equation}
This expression makes explicit the nature of shrinkage applied to $\hat{b}$ through the interaction of the global scale $\tau$, the local scales $\lambda$, and the structure-inducing matrix $\Omega$. To assess the specific influence of $\Omega$, we consider the difference between the posterior means obtained under $\Omega$ and the identity matrix $I$:
\begin{equation}
\label{eq:dif-cond-means}
\mathbb{E}(b \mid y, \lambda, \tau, \sigma, \Omega) - \mathbb{E}(b \mid y, \lambda, \tau, \sigma, I) = \left(Q_\Omega^{-1} - Q_I^{-1} \right) X' y.
\end{equation}
This formulation allows us to isolate and quantify the effect of introducing $\Omega$ on the regularization applied to the MLE. In particular, the term $Q_\Omega^{-1} - Q_I^{-1}$ captures the deviation from standard isotropic shrinkage, revealing how structured dependence alters the posterior behavior.

Let $A$, $B$, and $C$ be invertible matrices of dimension $p$, and let $x \in \mathbb{R}^p$ be a vector. Denote by $\|A\|_2$ the spectral norm of $A$, i.e., its largest singular value. To derive two-sided bounds for the spectral norm difference between $Q_\Omega^{-1}$ and $Q_I^{-1}$, we make use of the resolvent identity $A^{-1} - B^{-1} = A^{-1}(B - A)B^{-1}$, the submultiplicative property $\|AB\|_2 \leq \|A\|_2 \|B\|_2$ and its reverse form $\|ABC\|_2 \geq \|A^{-1}\|_2^{-1} \|B\|_2 \|C^{-1}\|_2^{-1}$ for invertible $A$ and $C$, as well as Weyl’s inequality for eigenvalues of symmetric matrices \citep{weyl_asymptotische_1912, horn_matrix_2012}. (See Appendix \ref{appendix} for the proof).

\begin{equation}
\label{eq:scaled-norm-bound}
\frac{\| \Omega^{-1} - I \|_2}{\lambda_1^2 
\left( \nu_1 + \frac{1}{\lambda_p^2 \omega_p} \right) 
\left( \nu_1 + \frac{1}{\lambda_p^2} \right)}
\leq 
\| Q_\Omega^{-1} - Q_I^{-1} \|_2   
\leq 
\frac{ \| \Omega^{-1} - I \|_2}{\lambda_p^2 
\left( \nu_p + \frac{1}{\lambda_1^2 \omega_1} \right) 
\left( \nu_p + \frac{1}{\lambda_1^2} \right)}.
\end{equation}

Here, $\lambda_1$, $\omega_1$, and $\nu_1$ denote the largest local scale, the largest eigenvalue of $\Omega$, and the largest eigenvalue of $X^\top X$, respectively, while $\lambda_p$, $\omega_p$, and $\nu_p$ denote the corresponding smallest values. We assume $\tau^2 = 1$ for notational simplicity. These inequalities quantify how structural deviations in the prior covariance, the spread of local shrinkage parameters, and the conditioning of the design matrix influence the posterior precision matrix. 

In particular, the difference between the structured and independent posterior precisions is amplified when shrinkage is globally strong (i.e. small $\lambda_1$ and hence small $\lambda_p$), the design is poorly conditioned (i.e. small $\nu_p$), or the prior structure deviates substantially from independence (i.e. large $\| \Omega^{-1} - I \|_2$). Conversely, the upper bound shows that the difference can be negligible when shrinkage is weak (i.e., large $\lambda_p$ and hence large $\lambda_1$), the design is well-conditioned (i.e. large $\nu_p$), and the structural deviation is modest (i.e. small $\| \Omega^{-1} - I \|_2$). In this regime, the dependence structure plays a limited role, and the posterior mean is largely data driven.

Crucially, $\Omega$ alters the geometry of the posterior. Through the matrix $D_\lambda \Omega D_\lambda$ in \eqref{eq:pc-mean-mle}, it interacts with the design matrix to modulate how prior structure shapes the shrinkage pattern. When $\Omega$ encodes interpretable structure, such as sparsity or neighborhood dependencies, it reweights the contribution of each component of $\hat{b}$ along meaningful axes in parameter space. Clearly, this can also enhance stability in settings with ill-conditioned designs or multicollinearity.

\subsubsection{Divergences}
\label{subsec:divergences}

We analyze how the inclusion of $\Omega$ alters our model by studying the divergence between probability measures, while keeping all other components fixed. In particular, we focus on the Kullback–Leibler (KL) divergence \citep{kullback_information_1951}, which quantifies the discrepancy between two probability distributions. A key advantage of using divergence-based metrics \citep{renyi_measures_1961} is that they account for the entire distribution, rather than relying solely on pointwise comparisons.

Let $\mathcal{P} \sim \normal( \mu_\mathcal{P} , \Sigma_\mathcal{P} )$ and $Q \sim \normal( \mu_\mathcal{Q} , \Sigma_\mathcal{Q} )$ be two $p$ dimensional multivariate normal distributions. Their KL divergence has the following closed-form expression \citep{pardo_statistical_2018}:
\begin{equation}
    \KL{\mathcal{P}}{\mathcal{Q}} = \frac{1}{2} \left[  (\mu_\mathcal{Q}- \mu_\mathcal{P})' \Sigma_\mathcal{Q}^{-1} (\mu_\mathcal{Q}- \mu_\mathcal{P}) + \text{tr} \left(\Sigma_\mathcal{Q}^{-1} \Sigma_\mathcal{P} \right) - \ln \frac{ |\Sigma_\mathcal{P}|  }{|\Sigma_\mathcal{Q}|} - p\right].
\end{equation}
Now consider $\mathcal{P}$ as the standard shrinkage prior for $b$ with $\Omega = I$ (see \eqref{eq:spprior}), and let $\mathcal{Q}$ be a prior of the form \eqref{eq:dasp-prior}. The Kullback–Leibler (KL) divergence between the conditional prior distributions for $b \mid \sigma, \tau, \lambda$ depends explicitly on $\Omega$ and is given by
\begin{equation}
\label{eq:KLpriors}
\KL{\mathcal{P}}{\mathcal{Q}} = \text{tr}\left(\Omega^{-1} \right)+ \ln |\Omega| - p.
\end{equation}
This expression emphasizes how  $\Omega$  influences the relative entropy between the two priors. The trace term  \text{tr}($\Omega^{-1}$)  indicates that nearly ill-conditioned (or nearly singular) matrices lead to a large KL divergence, reflecting different patterns of joint shrinkage. Specifically, this suggests that small eigenvalues of $\Omega$ result in strong heterogeneous shrinkage along particular directions, as the model is more sensitive to those directions with weaker or nearly non-invertible components. The term $\ln |\Omega|$  captures the overall volume scaling of the covariance structure imposed by  $\Omega$. The KL divergence of the conditional posteriors of $b \mid y, \sigma, \tau, \lambda$ also possesses a closed-form expression. We do not include it here since it is more difficult to interpret.

\begin{figure}[t]  
    \centering
    \includegraphics[width=\textwidth]{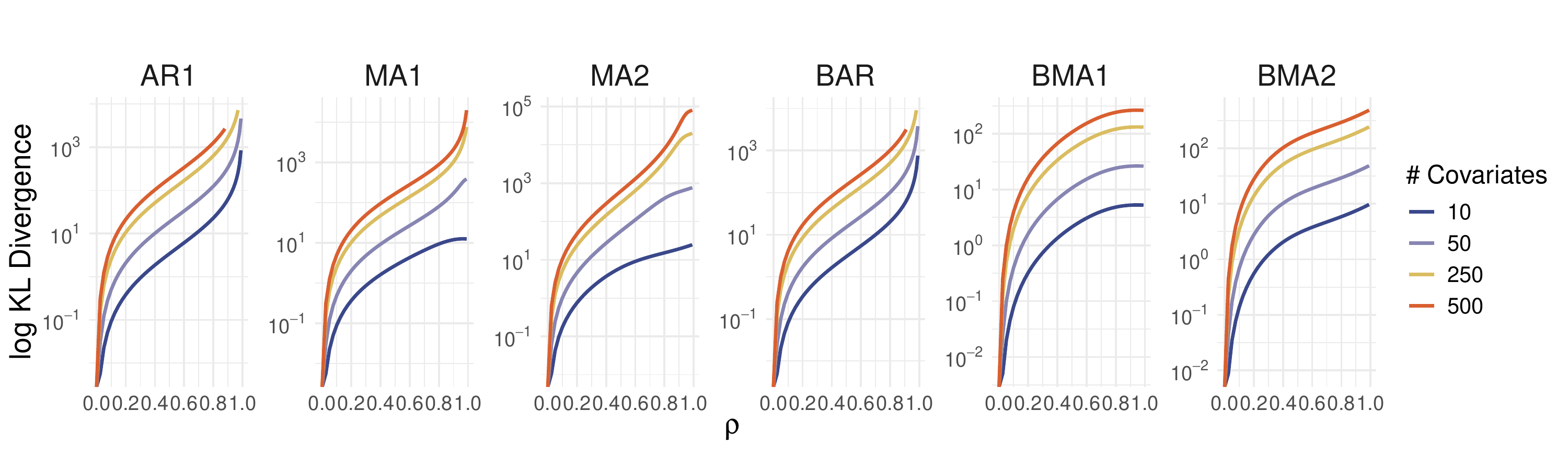}  
    \caption{ KL divergence as a function of the correlation parameter $\rho$ for various structures of $\Omega$. The corresponding algebraic forms are provided in Appendix~\ref{appendix}. }  
    \label{fig:prior-kldiv}  
\end{figure}

Figure \ref{fig:prior-kldiv} illustrates how the KL divergence varies across different specifications of $\Omega$. In the provided examples, each correlation matrix $\Omega$ is parameterized by a single hyperparameter $\rho \in (0,1)$, which fully determines its structure. We consider commonly used correlation patterns, including the autoregressive model of order 1 (AR1), moving average models of orders 1 and 2 (MA1 and MA2), as well as their blocked counterparts: BAR1, BMA1, and BMA2. These same structures are also employed in our experiments in Section \ref{sec:experiments}. We provide their algebraic definitions in Appendix \ref{appendix}.

The results in Figure \ref{fig:prior-kldiv} show that the KL divergence increases monotonically with the number of covariates, and the discrepancy between the conditional distributions becomes more pronounced as $\rho$ increases—especially with a sharp rise around $\rho \approx 0.9$. This indicates that differences between prior specifications are most substantial when parameters are strongly correlated. Consequently, we conjecture that posterior inference under the DASP will diverge most noticeably from standard shrinkage priors in high-correlation settings.

\subsubsection{Contour plots of marginal distributions}
\label{subsec:contour_plots}

The properties of shrinkage priors are typically analyzed through the marginal distributions of individual coefficients, $p(b_i)$, focusing on two desirable features: strong concentration near zero to shrink noise and heavy tails to retain large signals \citep{pas_horseshoe_2014, pas_conditions_2016, van_der_pas_theoretical_2021}. In contrast, the joint prior distribution of the full coefficient vector,
\begin{equation}
\label{eq:marginalb}
p(b) = \int p(b, \lambda, \tau) \, d\lambda \, d\tau,
\end{equation}
is rarely studied, largely because most priors assume conditional independence among the components of $b$. However, the joint distribution can reveal important aspects of the prior’s global behavior \citep{piironen_projective_2020}. In our setting, where we explicitly introduce dependencies across coefficients, examining $p(b)$ becomes particularly meaningful. Unlike the marginals, the joint captures how shrinkage is applied collectively across $b$, offering insights into how the prior balances local adaptivity with global structure.

\begin{figure}[t!]  
    \centering
    \includegraphics[width=\textwidth]{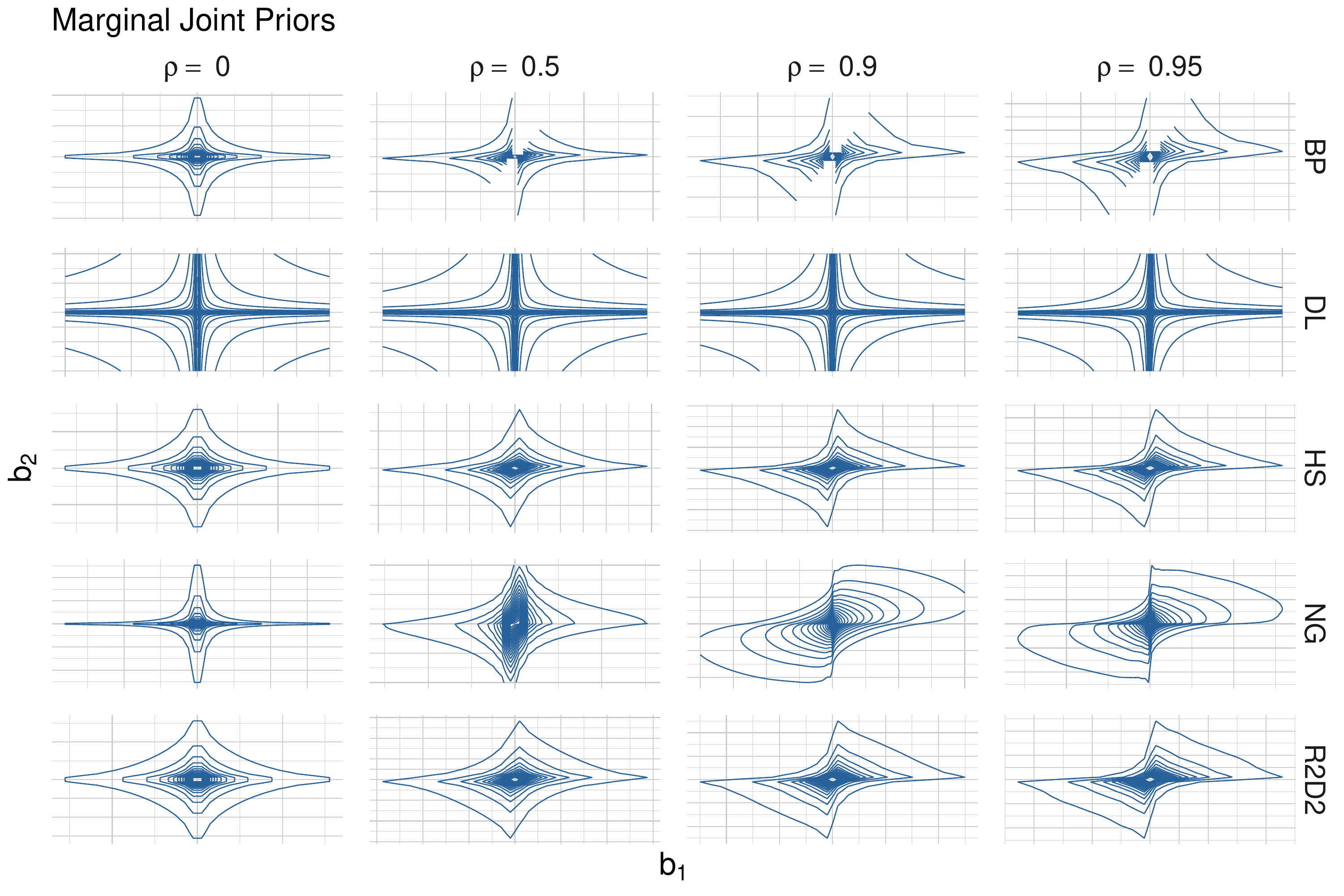}  
    \caption{Monte Carlo approximations of the bivariate joint marginal prior distribution $p(b_1, b_2)$ under different shrinkage priors:Beta Prime (BP),  Dirichlet-Laplace (DL), Horseshoe (HS), Normal-Gamma (NG), and R2D2, and for various correlation levels $\rho$.  We use default hyperparameters for each prior, with $\sigma = 1$, and $\Omega = (1 - \rho)I + \rho JJ'$, where $J$ is a $2$-dimensional vector of ones.}  
    \label{fig:prior-log}  
\end{figure}

Since a closed-form expression for $p(b)$ is generally intractable, we rely on Monte Carlo approximations \citep{robert_monte_2004}. Figure \ref{fig:prior-log} illustrates how introducing a dependency structure through $\Omega$ influences the joint prior distribution $p(b)$. In this analysis, we fix $\sigma = 1, p = 2$ and set $\Omega = \begin{pmatrix} 1 & \rho \\ \rho & 1 \end{pmatrix}$, varying the value of $\rho$ to control the degree of correlation. We apply this procedure to several widely used shrinkage priors: Beta Prime (BP), Dirichlet–Laplace (DL), Horseshoe (HS), Normal–Gamma (NG), and R2D2 \citep{bai_beta_2021,bhattacharya_dirichletlaplace_2015, carvalho_horseshoe_2010, griffin_inference_2010, zhang_bayesian_2020}, using default hyperparameter values as recommended in the literature. We show the form of these priors as well as their default hyperparameters in Appendix \ref{appendix}.

$\Omega$ represents the correlation between $b = (b_1, b_2)'$, as expressed in the conditional distribution $p(b \mid \lambda, \tau)$. However, the joint distribution $p(b)$ can exhibit a markedly different dependency structure. This is particularly evident because $p(b)$ is rarely a multivariate Gaussian, even when $p(b \mid \lambda, \tau)$ is. The plots in Figure \ref{fig:prior-log} show how the correlation from the conditional distribution of $b$ is propagated to the marginal distributions.

The standard uncorrelated priors are represented when $\rho = 0$. The contour plots in Figure \ref{fig:prior-log} show prior mass concentrated near the axes, indicating a prior preference for sparse vectors. For shrinkage to take place, it is expected that the contours show a non-convex behavior, approximating the axes. As $\rho$ increases, the prior mass shifts toward the main diagonal, suggesting that, a priori, the coefficients are expected to take similar values. the resulting convexity of certain contours indicates that the dependency-aware shrinkage prior may not always shrink for fixed values of  $\rho$, potentially weakening the prior’s regularizing effect. When high correlations are present, DASP forces both coefficients to take the same value. This behavior is akin to Ridge regularization, which also sets coefficients to the same values if they belong to highly correlated, same-scaled variables \citep{hoerl_ridge_1970}.

\subsubsection{Conditional distributions}
\begin{figure}[t]  
    \centering
    \includegraphics[width=\textwidth]{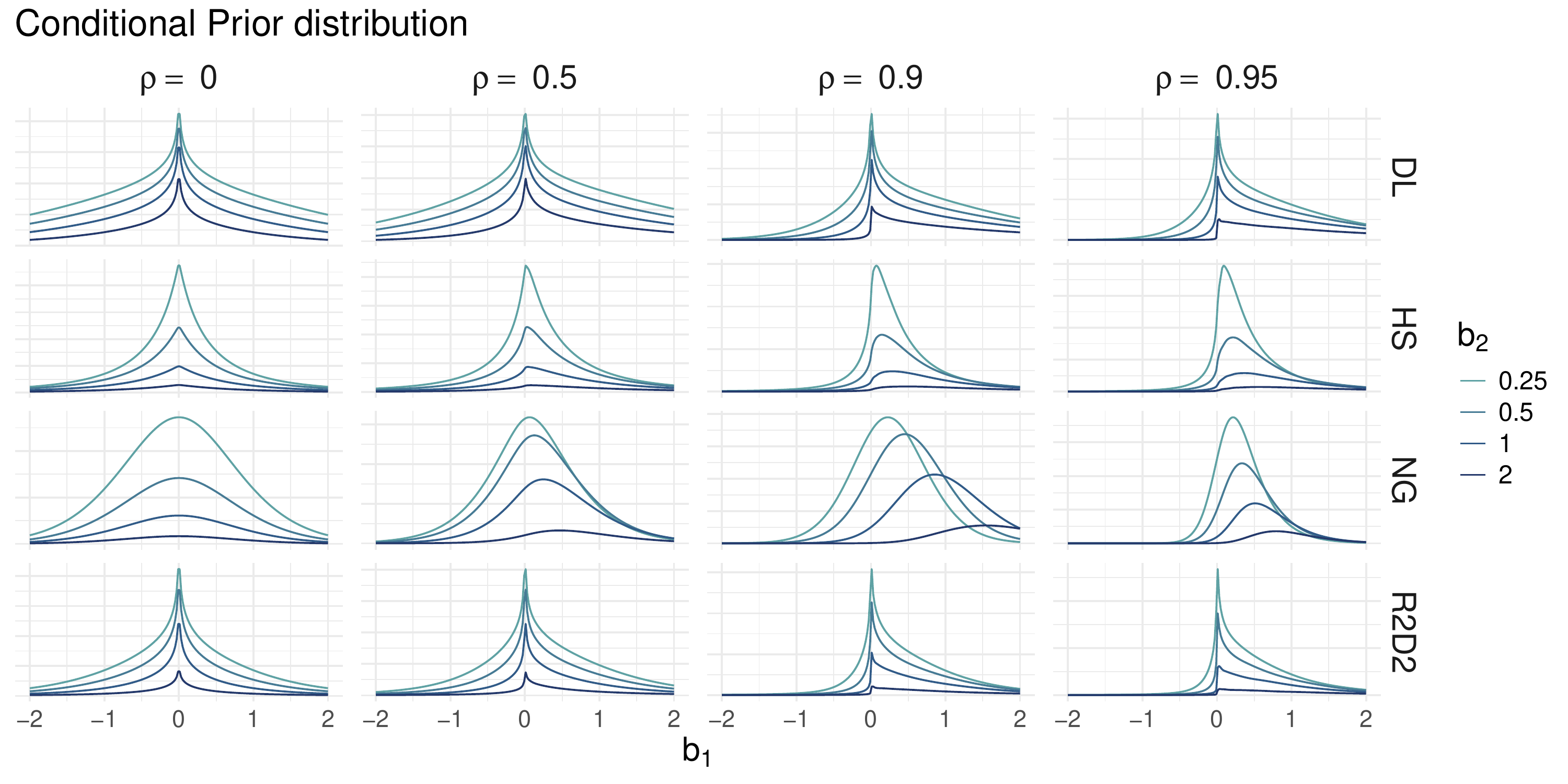}  
    \caption{Monte Carlo approximations of the unnormalized conditional prior distribution $p(b_1 \mid b_2)$ for varying values of $b_2$ and $\rho$. As both $b_2$ and $\rho$ increase, the distribution shifts toward larger values of $b_1$, indicating reduced shrinkage. 
 }  
    \label{fig:prior-cond}  
\end{figure}

Figure \ref{fig:prior-cond} displays Monte Carlo approximations of the conditional distributions $p(b_1 \mid b_2)$, using the same covariance structure $\Omega$ as in previous figures, while varying $\rho$ and $b_2$. When $\rho = 0$, increasing $b_2$ leads to a marginal distribution for $b_1$ with reduced mass near zero. As both $\rho$ and $b_2$ increase, the conditional distribution shifts markedly to the right, with $b_2$ primarily determining the magnitude of this shift. The pronounced spikes in the plot arise because we are visualizing $p(b_1 \mid b_2)$ rather than the fully conditional $p(b_1 \mid b_2, \lambda, \tau)$, which are Gaussian. Consistent with the contour plots in Figure \ref{fig:prior-log}, the marginal distribution becomes smoother as $\rho$ increases, inducing less shrinkage on $b_1$ in the presence of stronger correlation.

\subsubsection{Shrinkage properties}
\label{subsec:shrink_prop}

Shrinkage properties can be studied through shrinkage factors, which quantify how strongly each coefficient $b_i$ is pulled toward zero \citep{carvalho_horseshoe_2010, piironen_sparsity_2017, aguilar_intuitive_2023}. The well-known normal means problem \citep{castillo_needles_2012} arises in Model \eqref{eq:linregeq} when the design matrix is the identity, i.e., $X = I$. Under a standard shrinkage prior, an application of Fubini’s theorem yields $ \mathbb{E}(b_i \mid y_i, \lambda_i) = \frac{\lambda_i^2}{1+\lambda_i^2} = (1-\kappa_i) y_i$, where the quantity
\begin{equation}
\label{eq:kappa1}
 \kappa_i := \frac{1}{1+\lambda_i^2},
\end{equation}
is referred to as the shrinkage factor \citep{carvalho_horseshoe_2010, piironen_sparsity_2017, bai_large-scale_2019, aguilar_intuitive_2023}. It represents the proportion of the posterior mean of $b_i$ that is attributed to shrinkage toward zero after observing the data $y$. Since $0 \leq \kappa_i \leq 1$, the law of total expectation implies $ | \mbe(b_i | y_i) | = |(1- \mbe(\kappa_i | y_i)) y_i | \leq |y_i| $. In this context, $y_i$ serves as MLE of $b_i$, so $\kappa_i$ can be interpreted as the degree to which the MLE is shrunk toward zero. When $b_i$ corresponds to signal (i.e., a large effect), $\kappa_i \approx 0$, whereas for noise, $\kappa_i \approx 1$.

Shrinkage factors offer a unified framework for comparing the behavior of different shrinkage priors across a range of hyperparameter settings \citep{polson_local_2012, tadesse_handbook_2021}. 
A desirable property of such priors is their ability to effectively distinguish between noise and signal components. This can be studied by examining the prior distribution of the shrinkage factor $\kappa_i$, which reveals how aggressively the prior shrinks coefficients under varying hyperparameter configurations.

For instance, the Horseshoe prior gained popularity due to the bimodal nature of its implied distribution for $\kappa_i \sim \text{Beta}(1/2, 1/2)$, with mass concentrated near 0 and 1 \citep{carvalho_horseshoe_2010, armagan_generalized_2011, pas_uncertainty_2017}. This characteristic enables the prior to strongly shrink noise (large $\kappa_i$) while retaining signals (small $\kappa_i$). In contrast, earlier shrinkage priors often concentrated their mass either near minimal shrinkage or excessive shrinkage, limiting their adaptability \citep{carvalho_handling_2009}.

The notion of shrinkage factors from Equation \eqref{eq:kappa1} can be extended to settings with nontrivial dependency structures ($\Omega \neq I$) by examining the conditional posterior mean of $b \mid \sigma, \tau, \lambda $ given in Equation \eqref{eq:pc-var}. This allows us to assess how the posterior mean deviates from the MLE under more general dependence structures. Specifically, we define the matrix-valued shrinkage factor as:
\begin{equation}
\label{eq:kappa_omega}
    \kappa_\Omega = I -\tau^2 \, D_\lambda \Omega D_\lambda (\tau^2 \, D_\lambda \Omega D_\lambda+ (X'X)^{-1})^{-1}. \ \
\end{equation}
This definition becomes cumbersome when approaching shrinkage factors from the perspective of MLE shrinkage. Unlike the classical scalar shrinkage factors, $\kappa_\Omega$ is matrix-valued and generally non-diagonal, reflecting the induced dependence across coefficients due to $\Omega$. To illustrate this, consider the special case with $\tau = \sigma = 1$, $\Omega = \begin{pmatrix} 1 & \rho \\ \rho & 1 \end{pmatrix}$, and $X = I$. In this case, the conditional posterior mean becomes:
\begin{equation}
    \mbe(b \mid y, \lambda ) = \frac{1}{1+\lambda_1^2+ \lambda_2^2+ \lambda_1^2\lambda_2^2(1-\rho^2)} \begin{pmatrix}
        \lambda_1^2(1+ \lambda_2^2(1- \rho)) \hat{b}_1 +\rho \lambda_1 \lambda_2 \hat{b}_2 \\
        \lambda_2^2(1+ \lambda_1^2(1- \rho))\hat{b}_2 +\rho \lambda_1 \lambda_2 \hat{b}_1 \\
    \end{pmatrix}
\end{equation}
When $\rho = 0$, $\kappa_\Omega$ reduces to a diagonal matrix, and we recover the classical shrinkage factors $\kappa_i = \frac{1}{1 + \lambda_i^2}$ by taking the diagonal. The correlation parameter $\rho$ controls the degree of interaction between the local scales $\lambda_1$ and $\lambda_2$. As $\rho$ increases, the posterior means of $b_1$ and $b_2$ become more entangled, and shrinkage toward zero weakens unless the MLEs are already near zero. This also reflects a global pooling effect, where coefficients are increasingly pulled toward each other. Hence posterior estimates are pulled not only toward zero, but also toward one another. 

Importantly, the introduction of correlation through $\Omega$ does not eliminate shrinkage toward zero. Instead, it makes shrinkage less aggressive in the direction of differences between MLEs, especially when both estimates are large in magnitude. The correlation parameter $\rho$ controls a trade-off between promoting sparsity (via local shrinkage through $\lambda_i$) and encouraging similarity across coefficients (via pooling). Full shrinkage remains possible due to the continued flexibility of the local scales $\lambda_i$.

While shrinkage factors offer a local, coefficient-wise view of regularization, a more global perspective is provided by the effective number of parameters, $m_\text{eff}$, introduced by \cite{piironen_sparsity_2017} and defined as:
\begin{equation}
    m_\text{eff} = \sum_{i=1}^p( 1- \kappa_i).
\end{equation}
This quantity serves as a measure of the model’s effective dimensionality, satisfying $m_\text{eff} \leq p$, and reflects how the combination of local and global shrinkage affects model complexity.

When prior information or domain expertise suggests a plausible number of active predictors, visualizing the prior distribution of $m_\text{eff}$ under different hyperparameter settings offers practical guidance for prior elicitation \citep{piironen_sparsity_2017,aguilar_intuitive_2023}. The user can rely on $m_\text{eff}$ to match the expected model complexity to domain knowledge without needing to examine each $\lambda_i$ individually.  This approach is especially informative in high-dimensional regimes, where balancing flexibility and parsimony is critical. We extend this definition to our dependency-aware shrinkage prior by using the matrix valued shrinkage factor $\kappa_\Omega$, leading to
\begin{equation}
\label{eq:meff_dasp}
m_\text{eff} = \text{tr}(I - \kappa_\Omega).
\end{equation}

\begin{figure}[t]  
    \centering
    \includegraphics[width=\textwidth]{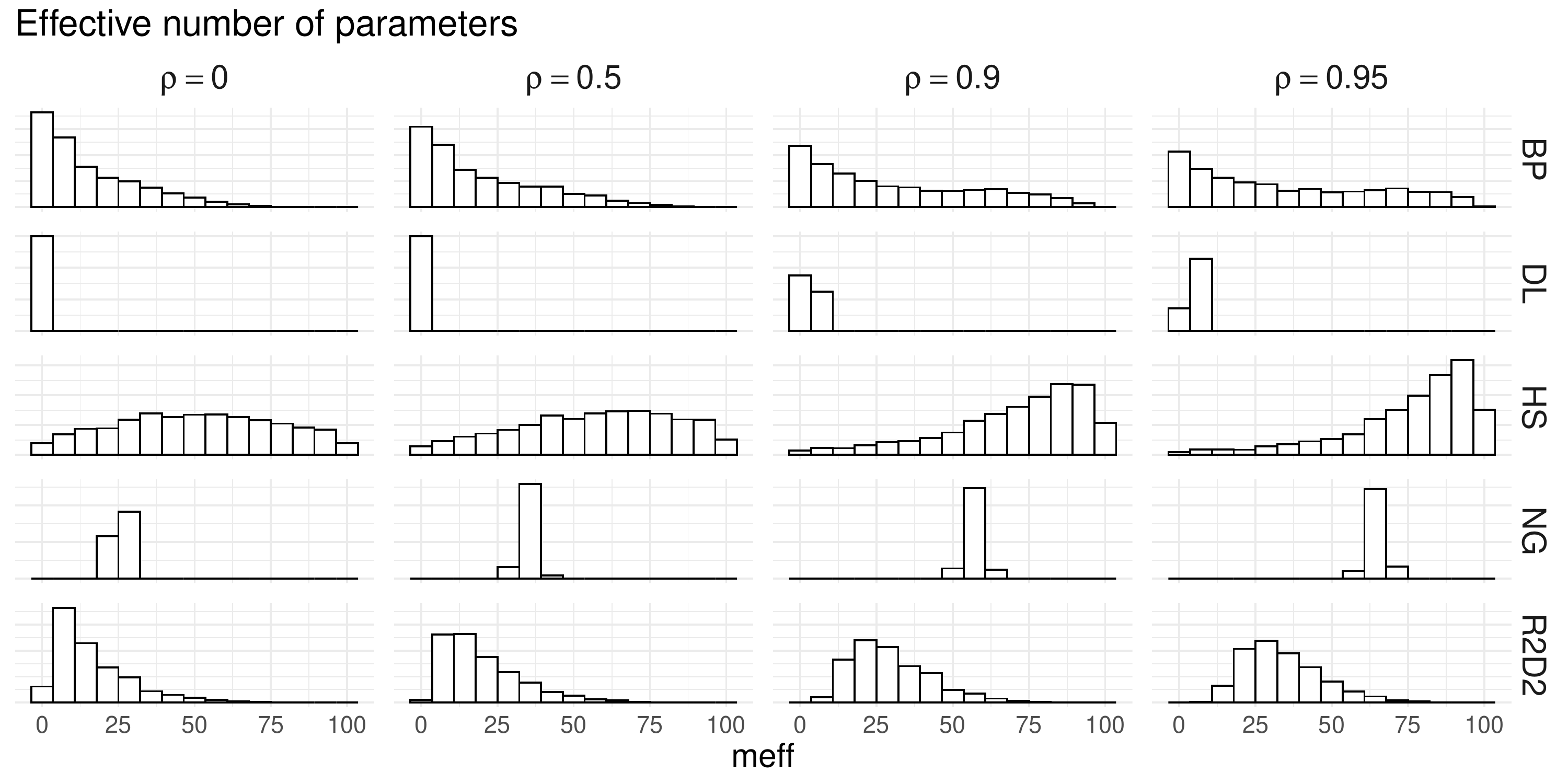}  
    \caption{Implied priors of the effective number of parameters (Equation \eqref{eq:kappa_omega}) for different shrinkage priors (rows) and correlations (columns). We have used the default hyperparameters for the different shrinkage priors, set $p=100$ covariates, $\sigma = 1$ and $\Omega = (1-\rho)I + \rho JJ'$, where $J$ is a $p$-dimensional vector of ones. }  
    \label{fig:prior-meff}  
\end{figure}

This extension preserves the original interpretation: the trace still quantifies the influence of the data on the posterior mean, but now incorporates the dependence structure among coefficients induced by $\Omega$. In other words, $m_\text{eff}$ continues to represent an effective model size, but under a prior where shrinkage is applied not just individually, but collectively. This generalization is particularly meaningful because it reflects how correlation alters the balance between sparsity (zero shrinkage) and similarity (pooling). Notably, we recover the classical definition when $\Omega = I$.

The posterior distribution of $m_\text{eff}$ serves as a useful diagnostic to monitor how the model adapts to the data. Comparing posterior expectations of $m_\text{eff}$ across different prior choices (e.g., independent vs. correlated shrinkage) helps quantify how strongly the priors influence model complexity in practice, and to what extent dependencies in $\Omega$ are effectively “used” in posterior inference.

We illustrate the prior distribution of $m_\text{eff}$ for various shrinkage priors in Figure \ref{fig:prior-meff}. For these simulations, we fixed $p = 100$, used $X = I$, and employed standard hyperparameters for each prior. The correlation structure was specified as $\Omega = (1-\rho)I + \rho JJ'$, where $J$ is a $p$-dimensional vector of ones, and we varied $\rho$ to assess the impact of correlation. As $\rho$ increases, the effective model size $m_\text{eff}$ grows, indicating less global shrinkage. This is due to the increased dependence among the coefficients, which results in more pooling across them. However, this effect can be moderated by adjusting the hyperparameters responsible for shrinkage, such as increasing concentration near zero, which would counterbalance the growth in effective model size induced by the correlation. 

\subsection{Specification of dependence structures}
\label{subsec:dep_structs}

A significant challenge in the use of model \eqref{eq:dasp-prior} is the specification of the matrix $\Omega$. We discuss our preferred approach to specifying $\Omega$. Potential alternative methods are detailed in the Discussion section.

It is possible to extract correlational information about the regression coefficients $b$ directly from the design matrix $X$. Since $X$ encodes information about how the predictors relate to each other, we can model the dependencies between the regression coefficients $b$ based on the structure of $X$. A well-known example of this approach is Zellner’s $g$ prior (see Section \ref{sec:other_priors}), which achieves this by setting the prior covariance matrix of the coefficients proportional to the covariance of MLE, i.e., $\operatorname{Cov}(b \mid \sigma, g) = g \,\sigma^2 (X^{\prime}X)^{-1}$ where $g > 0$.
An alternative interpretation of Zellner’s prior is that it uses a covariance structure for $b$ that is proportional to the true covariance matrix of the predictors, $\Sigma_X$. This is because when $n$ is sufficiently large, the unbiased estimator of the covariance of $X$ is approximately $X'X / n$, which converges to $\Sigma_X$ as $n \to \infty$ and $p$ is fixed.

Inspired by Zellner’s prior, we propose a generalization of shrinkage priors that incorporates the structure of the design matrix $X$, aligning with the prior structure of our proposed model in Equation \eqref{eq:dasp-prior}. Specifically, we derive the dependence structure $\Omega$ directly from $X$, using its covariance matrix $\Sigma_X$ to inform the relationships between the regression coefficients.

Let $\operatorname{Cov}(X)= \Sigma_X$ denote the true covariance matrix of $X$. To define $\Omega$, we first compute the inverse of $\Sigma_X$, i.e., the precision matrix of $X$, which we denote by $\Theta_X = \Sigma_X^{-1}$. We then standardize $\Theta_X$ to obtain a correlation matrix, which we use as $\Omega$. Explicitly, we write:
\begin{equation}
\label{eq:omega-sigmax}
\Omega = \operatorname{Cor}(\Theta_X) = \operatorname{Cor}\left(\Sigma_X^{-1}\right),
\end{equation}
where the $\operatorname{Cor}$ operator indicates standardization of the matrix by dividing each off-diagonal entry by the product of the corresponding diagonal entries. This results in a matrix $\Omega$ that encodes the correlation structure between the regression coefficients, reflecting the dependencies among them based on the design matrix $X$. This leads to the following prior covariance for $b$:
\begin{equation}
\label{eq:cov-sigmax}
\operatorname{Cov}(b \mid \lambda, \tau, \sigma) = \tau^2 D_\lambda \operatorname{Cor} \left( \Sigma_X^{-1} \right) D_\lambda.
\end{equation}
This choice of $\Omega$ has several key interpretations:

\begin{enumerate}
    \item \textbf{Relationship to the partial precisions}. Let $\Psi_X$ denote the partial precision matrix of $X$ \citep{lauritzen_graphical_1996,lam_high-dimensional_2020}. Under our formulation, we have
    \begin{equation*}
      \Omega_{ij} = -\Psi_{ij}, \quad i \neq j.  
    \end{equation*}
    which implies that $\Omega_{ij}$ encodes the conditional correlation between $b_i$ and $b_j$ after adjusting for all other variables. Since $b$ follows a conditional multivariate Gaussian distribution, whenever $\Omega_{ij} = 0$, the coefficients $b_i$ and $b_j$ are conditionally uncorrelated given all other variables \citep{giraud_introduction_2014}. This structure allows us to model the dependencies between the regression coefficients directly from the design matrix. The relationship to partial precision matrices emphasizes how $\Omega$ reflects conditional dependencies between coefficients, which is critical for understanding the underlying structure in the regression model.
    
    \item \textbf{Prior-Likelihood agreement}. 
    The conditional posterior mean of $b \mid \sigma, \tau, \lambda$ (see Equation \eqref{eq:pc-mean-mle}) suggests that aligning the prior correlation  structure of $\Omega$ with the correlation structure implied by $(X'X)^{-1}$ enhances the coherence between the prior and the likelihood \citep{zellner_models_1996,ohagan_bayesian_2012}. Specifically, both the prior and the likelihood would then imply the same posterior correlation structure. In principle, this alignment should reduce prior-likelihood conflict and facilitate inference. 

    \item \textbf{Imaginary observations}. Another interpretation of this prior structure is in terms of imaginary or pseudo-observations. Specifically, the model can be viewed as augmenting the data with additional pseudo observations that share the same design matrix $X$, but with response values set to zero \citep{zellner1986assessing}. These pseudo-observations are scaled by the coefficients $\lambda_i$ and $\tau$, which determine their relative weight. Smaller values of $\lambda_i^2 \tau^2$ correspond to fewer pseudo-observations for the $i$th coefficient, resulting in a weaker prior, while larger values strengthen the prior by incorporating more pseudo-data. This perspective emphasizes how the prior adjusts based on the observed data. 

\end{enumerate}

In practice, the true covariance matrix $\Sigma_X$ is typically unknown. Instead, it must be estimated from the observed design matrix $X$, which can be particularly challenging when $p > n$. We discuss methods for estimating $\Sigma_X$ in Section \ref{sec:estimators_cov_matrices}. By using a suitable estimator for $\Sigma_X$, we can proceed with standard Bayesian inference. Importantly, since the entire regression model is conditioned on $X$, allowing the prior to depend on it poses no issues in terms of consistency or model specification .

\subsection{Estimation of covariance matrices}
\label{sec:estimators_cov_matrices}

In practice, $\Sigma_X$ is rarely known and must be estimated from the design matrix $X$. When $p < n$, and $X$ is of full rank, a natural approach is to proceed in a way that mirrors Zellner’s $g$ prior. Specifically, we propose setting the correlation structure implied by $X$ as:
\begin{equation}
\label{eq:omega-ngreaterp}
\Omega = \operatorname{Cor} \left( S^{-1} \right),
\end{equation}
where $S = \frac{1}{n-1} X' X$ is the sample estimator of $\Sigma_X$, which is an unbiased estimator of $\Sigma_X$, i.e., $\mathbb{E}(S) = \Sigma_X$ \citep{anderson_introduction_2003}. This choice assumes that the design matrix $X$ contains sufficient information to estimate the relationships between the predictors and therefore the dependence structure of the regression coefficients \citep{zellner1986assessing, george_variable_1993}.

When $X$ is not of full rank or when $p > n$, a different approach is required. To illustrate why, consider the sample covariance estimator $S$ of $\Sigma_X$. It is known that this estimator performs poorly when $p$ is large relative to $n$. In particular, when $p/n \to c \in (0, \infty)$ as both $p$ and $n$ grow, $S$ becomes an inconsistent estimator of $\Sigma_X$  \citep{ledoit_honey_2004, ledoit_well-conditioned_2004, lam_high-dimensional_2020, oriol_ledoit-wolf_2025}. Even in the simple case where $\Sigma_X = I$, the empirical distribution of the eigenvalues of $S$ will not converge to a point mass at 1, as expected. Instead, it will follow the Marčenko-Pastur distribution \citep{marcenko_distribution_1967, bai_spectral_2010, wainwright_high-dimensional_2019}. This phenomenon highlights the need for a more robust approach to approximating $\Sigma_X$.

Among the wide range of methods proposed for estimating $\Sigma_X$ in high-dimensional settings, we focus on approaches that (i) do not impose structural assumptions for consistency guarantees and (ii) prevent the presence of diverging eigenvalues as $n, p \to \infty$. This ensures that our approach remains as general as possible while minimizing the burden on users, who would otherwise need to specify structural assumptions on $\Sigma_X$ or provide prior information about the data. Shrinkage covariance estimation, in particular, provides well-conditioned estimates and has demonstrated significant empirical improvements, even in complex settings \citep{lam_high-dimensional_2020, ledoit_analytical_2020}.

A key advancement in this area is the linear shrinkage estimator introduced by \citet{ledoit_well-conditioned_2004}, given by
\begin{equation}
\label{eq:ledoit}
\Sigma^* =  \varphi_1 I + \varphi_2 S = \frac{\beta^2}{ \delta^2} \mu I + \frac{\alpha^2}{\delta^2} {S}.
\end{equation}
where $\delta^2 = \alpha^2 + \beta^2$ and $\Sigma^*$ minimizes the expected quadratic loss $\mathbb{E} \left\| \tilde{\Sigma} - \Sigma_X \right\|^2_F$ subject to $\tilde{\Sigma}= \varphi_1 I + \varphi_2 S$, with respect to nonrandom coefficients $\varphi_1, \varphi_2$. Here, $\|\cdot \|_F$ denotes the scaled Frobenius norm, i.e., $\| A \|_F = \text{tr}(A'A)/p$. The shrinkage parameters in Equation \ref{eq:ledoit} are computed as follows:
\begin{equation}
\label{eq:ledoit_wolf_params}    
\mu = \frac{\text{tr}(S)}{p}, \quad \alpha^2 = \| \Sigma_X - \mu I \|_F^2, \quad \beta^2 = \mathbb{E} \| S - \Sigma_X \|^2_F.
\end{equation}

Since $\Sigma_X$ is unknown, consistent estimators for $\mu$, $\alpha^2$, and $\beta^2$ must be used to construct a consistent estimator $S^*$ of $\Sigma^*$. The explicit form of $S^*$ is provided in Appendix~\ref{appendix}; see also \citet{ledoit_well-conditioned_2004} for derivation details.

One key advantage of the LedoitWolf $S^*$ estimator is that it does not require computationally expensive procedures, such as cross-validation or numerical optimization. Moreover, \cite{ledoit_well-conditioned_2004} demonstrate that its empirical counterpart $S^*$ is consistent under high dimensional general asymptotics-where both $p$ and $n$ grow at a proportional rate \citep{girko_g-analysis_1992, silverstein_strong_1995}- retaining the same properties as $\Sigma^*$. Furthermore, this estimator remains optimal in terms of minimizing the expected Frobenius loss, regardless of the distribution of the design matrix $X$ \citep{ledoit_well-conditioned_2004, ledoit_analytical_2020, lam_high-dimensional_2020}.

The resulting covariance matrix is always positive definite and well-conditioned, even in settings where $p \gg n$. Crucially, $S^*$ preserves the eigenvectors of the sample covariance matrix $S$ while shrinking the eigenvalues toward a multiple of the identity matrix \citep{ledoit_honey_2004,ledoit_well-conditioned_2004}. Although there are extensions of this method—such as shrinkage toward alternative target matrices or combining linear shrinkage with sparsity constraints—these typically introduce additional hyperparameters that require careful tuning \citep{ledoit_analytical_2020, lam_high-dimensional_2020, oriol_ledoit-wolf_2025}. Given our goal of minimizing the user’s burden, we focus on the standard Ledoit-Wolf estimator. Consequently, we practically compute $\Omega$ as
\begin{equation}
\label{eq:omega-nlessp}
\Omega = \operatorname{Cor} \left( (S^{*})^{-1} \right).
\end{equation}

\subsection{Workflow of the Method}
\label{subsec:workflow}

We shortly summarize the practical steps for applying a dependency-aware shrinkage prior. The procedure begins by selecting a standard shrinkage prior. If the covariance matrix $\Sigma_X$ is known or specified by the user, the dependence matrix $\Omega$ is computed directly via Equation~\eqref{eq:omega-sigmax}. When $\Sigma_X$ is unknown and $p < n$, $\Omega$ is constructed using the sample covariance matrix as per Equation~\eqref{eq:omega-ngreaterp}. In high-dimensional settings ($p > n$) or when $X$ is ill-conditioned, we recommend estimating $\Sigma_X$ using the Ledoit–Wolf shrinkage estimator $S^*$, Equation~\eqref{eq:ledoit}, and computing $\Omega$ via Equation~\eqref{eq:omega-nlessp}. This workflow integrates dependence information into the prior while maintaining numerical stability. We adopt this procedure in the experiments presented in Section~\ref{sec:experiments}.

\section{Experiments}
\label{sec:experiments}







\subsection{Simulations}
\label{sec:sims}

We conducted simulation studies to evaluate the impact of incorporating dependency structures in shrinkage priors on model performance. Specifically, we examined whether encoding dependence via correlation matrices, as discussed in Section \ref{sec:methods}, influences parameter recovery and predictive performance. Our simulation conditions reflect realistic scenarios encountered in practice. Additionally, we analyze real world datasets commonly used as benchmarks for dependence-structured models. This allows us to compare commonly observed empirical covariance structures with those assumed in our simulations.  

We considered the following shrinkage priors: Beta Prime (BP), Dirichlet-Laplace (DL), Horseshoe (HS), Regularized Horseshoe (RHS), and R2D2 (D2) \citep{bai_large-scale_2019, carvalho_horseshoe_2010, bhattacharya_dirichletlaplace_2015, piironen_sparsity_2017, zhang_bayesian_2020}. These priors have been widely adopted in Bayesian modeling over the last decade due to their strong theoretical properties and empirical performance. While this list is not exhaustive, it adequately represents the main approaches explored in shrinkage prior research. Given that we are interested in the sparse scenario, we have used default hyperparameters that imply strong shrinkage as recommended in the literature. See Appendix \ref{appendix} for details. 

All models were implemented in Stan \citep{carpenter_stan_2017, stan_development_team_stan_2024}, which employs an adaptive Hamiltonian Monte Carlo (HMC) sampler known as the No-U-Turn Sampler (NUTS) \citep{neal_mcmc_2011, brooks_handbook_2011, hoffman_no-u-turn_2014} to sample draws from posterior distributions. The associated data and code are available at \myosfresults.  

\subsubsection{Evaluation Metrics}  
\label{sec:eval_metrics}

To assess the effect of incorporating correlation structures into shrinkage priors, we compare models with and without dependence structures. Let $M$ denote a model with a standard shrinkage prior and $M_{\Omega}$ the same model incorporating a correlation matrix $\Omega$. Since the only difference between these models is the inclusion of $\Omega$, any variation in performance can be attributed to its presence. For a given quantity of interest $\mathcal{Q}$, we define the change in performance as $\Delta \mathcal{Q} = \mathcal{Q}(M_{\Omega}) - \mathcal{Q}(M).$
We evaluate model performance using two primary metrics: out-of-sample predictive performance and parameter recovery \citep{robert_bayesian_2007, vehtari_survey_2012}.  

\paragraph{Out-of-Sample Predictive Performance}  

Predictive accuracy was assessed using the expected log-pointwise predictive density ($\elpd$) \citep{vehtari_survey_2012, vehtari_practical_2016}, computed as $\elpd = \sum_{i=1}^{N_{\rm new}}\ln \left( \frac{1}{S} \sum_{s=1}^S p(y_{\text{new},i} \mid \theta^{(s)}) \right),$
where $\theta^{(s)}$ represents the $s$th draw from the posterior distribution $p( \theta \mid y)$ for $s = 1, \dots, S$. The $\elpd$ quantifies predictive performance across a set of $N_{\rm new}$ observations unseen during model training. Higher $\elpd$ values indicating better accuracy \citep{bernardo_bayesian_2009,vehtari_survey_2012}. Since log-probability scores are widely recommended as a default Bayesian evaluation metric, $\elpd$ provides a robust benchmark for model comparison \citep{bernardo_bayesian_2009,vehtari_practical_2016}.  

\paragraph{Parameter Recovery} We assess parameter recovery by computing the posterior Root Mean Squared Error (RMSE) for the regression coefficients $b$, defined as $\rmse = \frac{1}{K} \sum_{k=1}^K \sqrt{\frac{1}{S} \sum_{s=1}^S \left(b_k^{(s)} - b_k \right)^2}$,
where $b_k^{(s)}$ is the $s$th posterior draw for coefficient $b_k$, and $b_k$ denotes its true value. RMSE provides a global measure of estimation error, naturally capturing the bias-variance tradeoff \citep{robert_bayesian_2007,bernardo_bayesian_2009}. To gain deeper insights, we compute three variations of RMSE:   1) Overall RMSE (averaged across all coefficients). 2) RMSE for truly zero coefficients (measuring the ability to shrink the effect of irrelevant predictors). 3) RMSE for truly nonzero coefficients (assessing how well true signals are recovered).  

\paragraph{Coverage} We examine coverage properties by using $95\%$ marginal credibility intervals. We report average coverage proportion, average width, sensitivity, specificity (power) and coverage of non zeros \citep{neyman_ix_1997, berger_statistical_1985, benjamini_controlling_1995}. We also show Receiver operating characteristic (ROC) curves to understand how coverage properties would change as we modifiy the size of the credibility interval \citep{bradley_roc_1997}. See Appendix \ref{appendix} for these results.
 
\paragraph{Convergence Diagnostics} To ensure reliable posterior inference, we report two key MCMC diagnostics:  
1) The $\hat{R}$ statistic \citep{vehtari_rank_normalization_2021}, which compares within-chain and between-chain variance with values close to 1.0 indicating convergence. 2) The Effective Sample Size (ESS) \citep{geyer_practical_1992}, which measures the number of effectively independent samples. Higher values indicate lower autocorrelation and better mixing \citep{brooks_handbook_2011, margossian_nested_2024}.

\subsubsection{Generative Models}

We use model \eqref{eq:linregeq} as the generative model for our simulations, adapting it to accommodate various types of data encountered in practice, varying sparsity levels and correlation structures among covariates. The design matrix $X$ is drawn from a zero-mean multivariate normal distribution with covariance $\Sigma_X$, generated from one of the following: (1) AR1: autoregressive of order 1; (2) MA$q$ : moving average processes with $q \in {1, 2}$; and (3) BAR1 and BMA$q$: blocked versions using nonoverlapping blocks of size 5 with zero off-block correlations. All structures are parameterized by $\rho \in {0.5, 0.95}$, which governs the correlation strength. See Appendix~\ref{appendix} for details.
 
We fix $n = 100$ while varying $p \in \{50, 250\}$ to study both low and high dimensional settings. The residual variance $\sigma^2$ is calibrated to yield $R^2_0 \in \{0.5, 0.8\}$ \citep{gelman_bayesian_2013}. We construct $\Omega$ using either the true $\Sigma_X$ or its estimate: the sample covariance $S$ when $p < n$, and the Ledoit–Wolf shrinkage estimator $\tilde{S}$ when $p > n$.

Regression coefficients $b_i$ are generated under two schemes commonly used in the shrinkage prior literature \citep{carvalho_horseshoe_2010, griffin_inference_2010, bhattacharya_dirichletlaplace_2015, zhang_bayesian_2020}. 1) In the \textit{Block Coefficients} setup, the first and last five entries of $b$ are set to $b^* \in \{3, 7\}$, with all others set to zero. 2) In the \textit{Random Block Coefficients} setup, the same blocks are sampled from $\mathcal{N}(0, \Sigma_b)$, while the rest remain zero. We consider two versions of $\Sigma_b$: (A) diagonal with variance 9, and (B) AR(1) with $\rho_b = 0.8$ and the same variance. Sparsity is induced by setting each coefficient to zero with probability 0.75. Lastly, we sample the intercept as $b_0 \sim \mathcal{N}(0, 3)$.

\subsubsection{Results}

\begin{figure}[t]
    \centering
    \includegraphics[width=1\linewidth]{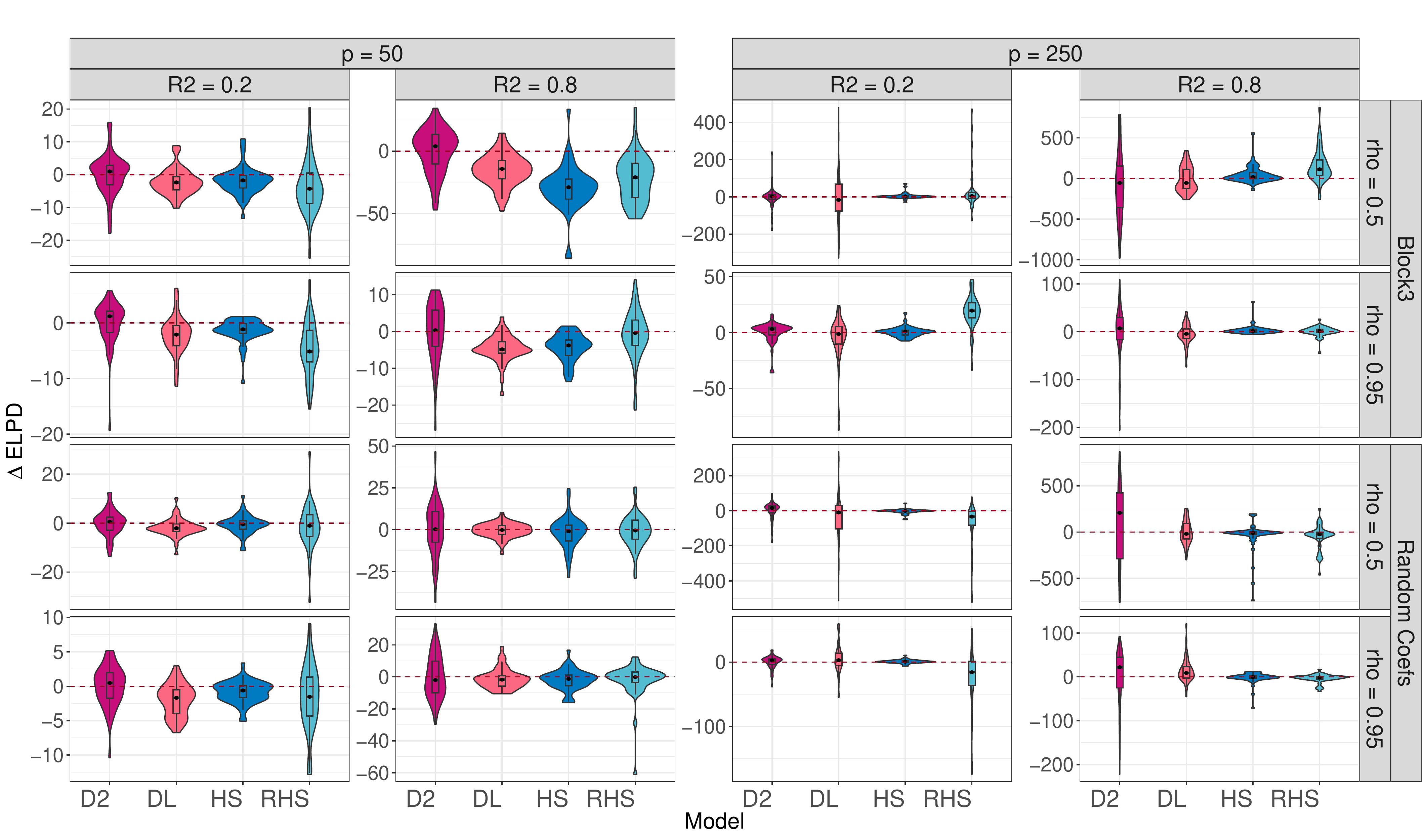}
    \vspace{-0.5cm}  
    \caption{ \textbf{ $\Delta$ELPD for each model under the BAR1 structure across simulation scenarios.} Model names represent the difference between the dependency-aware and standard versions, where positive values indicate improved predictive performance of the dependency-aware versions. We have omitted the BP prior due to high variability. } 
    \label{fig:lpd_test_blockedar}
\end{figure}

We focus on the subset of simulations using the blocked versions of the correlation matrix with the true value of $\Sigma_X$. The results presented here are representative of the full results, which can be found on OSF (\myosfresults). 

We hypothesize that our dependency-aware shrinkage priors are most effective when groups (or blocks) of highly correlated predictors collectively carry strong signal, i.e when several correlated covariates each contribute substantially to predicting the response.  In such cases, encoding correlation in the prior should allow for more coherent regularization across coefficients, improving parameter recovery under structured sparsity. However, a particularly challenging scenario arises when strong signal predictors are highly correlated with irrelevant predictors. Here, the prior faces a tension between two competing goals: promoting sparsity by shrinking noise coefficients to zero, and preserving signal strength by allowing large coefficients to escape shrinkage. This trade-off can lead to suboptimal inference—either by over-shrinking the signal due to its proximity to noise, or by inflating noise variables that inherit mass from nearby signals. 
\begin{figure}[t]
    \centering
    \includegraphics[width=1\linewidth]{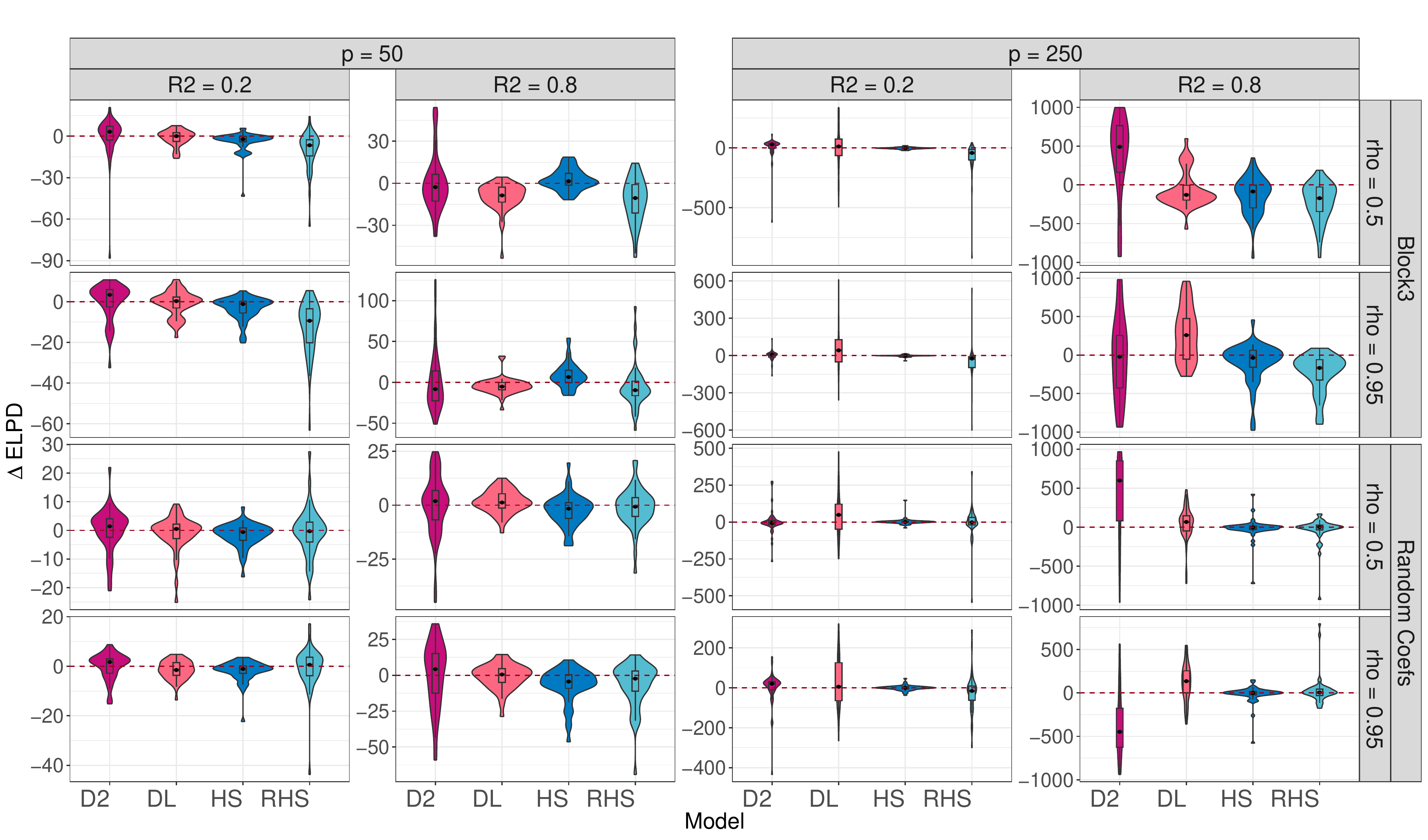}
    \vspace{-0.5cm}  
    \caption{ \textbf{ $\Delta$ELPD for each model under the BMA1 structure across simulation scenarios.} Model names represent the difference between the dependency-aware and standard versions, where positive values indicate improved predictive performance of the dependency-aware versions. We have omitted the BP prior due to high variability.} 
    \label{fig:lpd_test_blockedma2}
\end{figure}

Figures \ref{fig:lpd_test_blockedar} and \ref{fig:lpd_test_blockedma2} display the results of the BAR1 and BMA1 experiments, respectively, and illustrate the impact of incorporating the dependency matrix $\Omega$ on predictive performance. Overall, the findings indicate that predictive performance is not meaningfully improved by introducing dependency structures into shrinkage priors. In most configurations, the distribution of ELPD differences is centered near zero, suggesting that dependency-aware priors perform comparably to their standard counterparts, at least on average. This pattern is particularly evident in the low-dimensional setting, where even under high correlation and clearly structured signals, the inclusion of $\Omega$ offers no measurable advantage.

Similarly, in high dimensional settings, the inclusion of dependency structures does not produce strong predictive improvements across priors and design scenarios. These results imply that standard shrinkage priors are already capable of achieving strong predictive performance and that additional modeling flexibility through dependence structures does not translate into predictive gains. However, a notable observation is that predictive accuracy remains stable across all conditions, indicating that dependency-aware priors do not harm generalization. While gains in prediction are limited, these priors remain useful when the focus shifts to parameter recovery, which we discuss below.

\begin{figure}[t]
    \centering
    \includegraphics[width=1\linewidth]{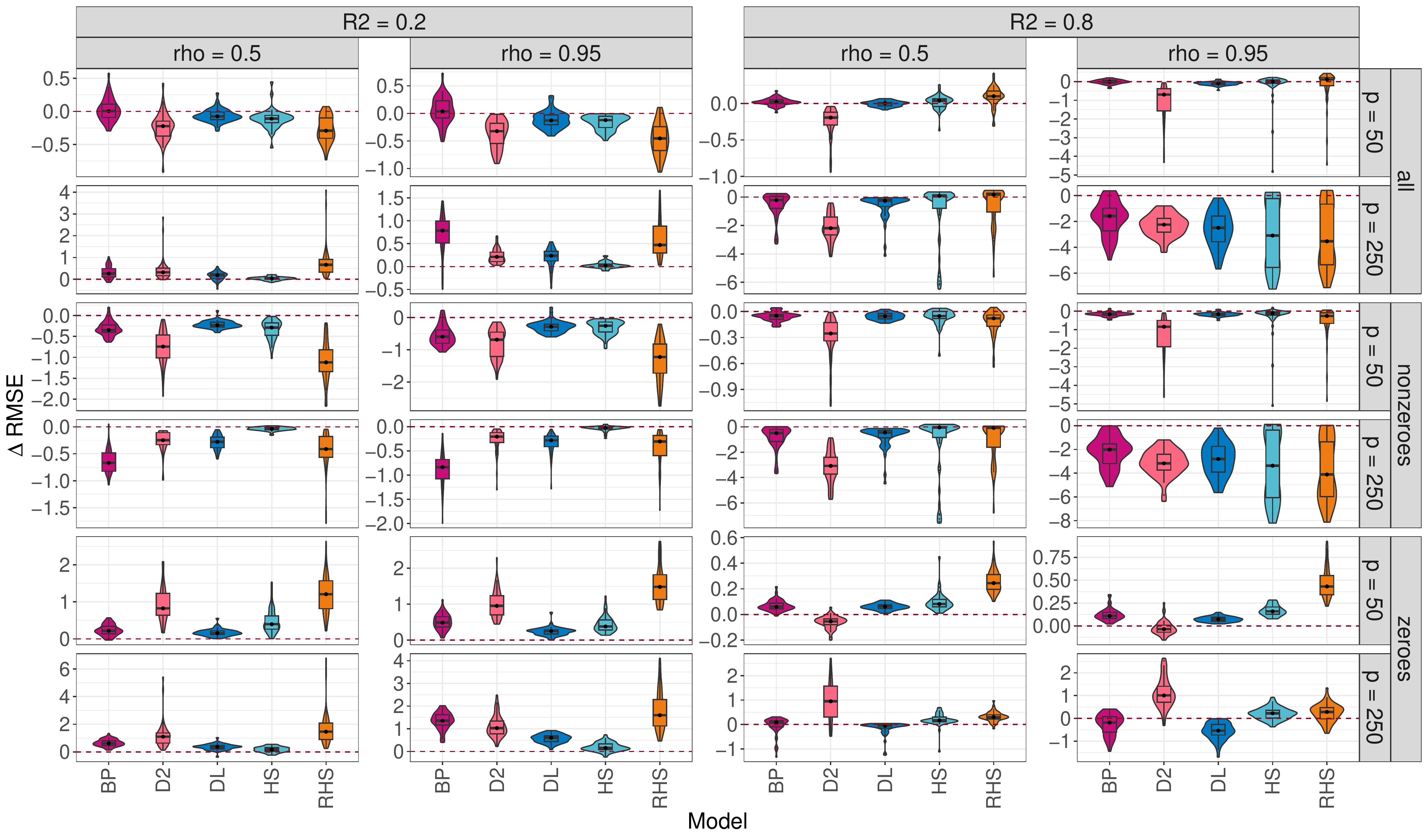}
    \vspace{-0.5cm}  
    \caption{ \textbf{ $\Delta$ RMSE for each model under the BMA1 structure with fixed block signals}. Rows correspond to all coefficients, nonzero (signal) coefficients, and zero (noise) coefficients. Model names represent the difference between the dependency-aware and standard versions, where negative values indicate improved parameter recovery of the dependency-aware versions. Dependency-aware priors show consistent gains for signal coefficients, while differences for noise coefficients are more modest. } 
    \label{fig:rmse_postb_cases_blockedma1}
\end{figure}

\begin{figure}[t]
    \centering
    \includegraphics[width=1\linewidth]{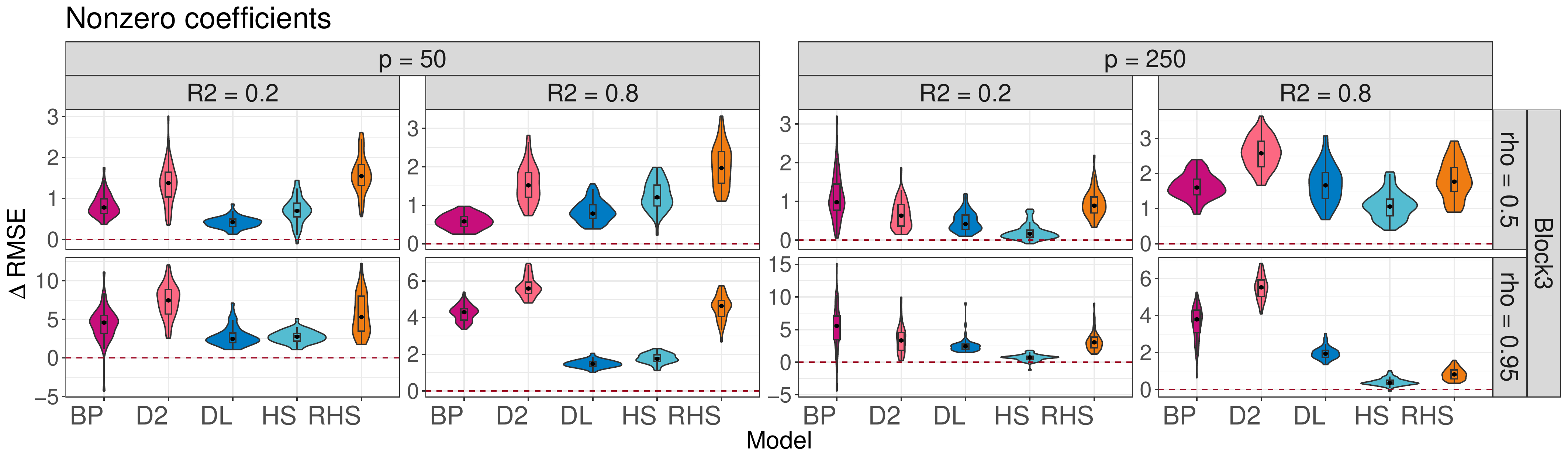}
    \vspace{-0.5cm}  
    \caption{  \textbf{ $\Delta$ RMSE for each model under the BMA1 structure with fixed block signals.} Only nonzero coefficients are shown. Model names represent the difference between the dependency-aware and standard versions, where negative values indicate improved parameter recovery of the dependency-aware versions. Improvements are limited and parameter recovery often worsens across conditions.} 
    \label{fig:rmse_postb_signal_blockedar}
\end{figure}

Figures \ref{fig:rmse_postb_cases_blockedma1} and \ref{fig:rmse_postb_signal_blockedar} illustrate the impact of incorporating correlational information into the prior on parameter recovery under the fixed block coefficient setting, for the BMA1 and BAR1 structures, respectively. Figure \ref{fig:rmse_postb_cases_blockedma1} shows that incorporating dependency information consistently reduces RMSE for nonzero coefficients across all scenarios. Notably, improvements appear even in the most challenging setting: high dimensionality ($p = 250$), low signal ($R^2 = 0.2$), and strong correlation ($\rho = 0.95$). When the signal is stronger ($R^2 = 0.8$), and thus noise is reduced, dependency-aware shrinkage priors still yield substantial improvements in parameter recovery. These results support the hypothesis that when $\Omega$ accurately reflects the groupwise structure of the relevant predictors, dependency-aware priors facilitate more effective signal recovery. Figure \ref{fig:rmse_postb_signal_blockedar} focuses on recovery of nonzero coefficients under the BAR1 structure. We limit the display to signal coefficients due to space constraints, but the results for all and zero coefficients mirror the patterns observed under BMA1. Contrary to BMA1, however, recovery of nonzero coefficients for BAR1 is not improved, and in fact, RMSE worsens uniformly across conditions.

The difference in performance between BMA1 and BAR1 can be attributed to the structure of the prior correlation matrix $\Omega$. While both scenarios assign strong signals to groups of correlated predictors, the implied $\Omega$ matrices differ in how they encode these relationships. In BAR1, $\Omega$ has a tridiagonal block structure, inducing local dependencies that connect only neighboring predictors within each block. In contrast, BMA1 produces blocks in which all predictors within a block are strongly correlated with one another, resulting in a dense (possibly high) correlation pattern. This richer within-block structure in $\Omega$ leads to stronger prior correlations among the nonzero coefficients, which enhances parameter recovery in the presence of structured signals. These results suggest that the effectiveness of dependency-aware priors depends not just on the presence of correlated signals, but on how exactly the signals are correlated.

\subsection{Real world case studies}

\begin{figure}[b]
    \centering
    \includegraphics[width=0.95\linewidth]{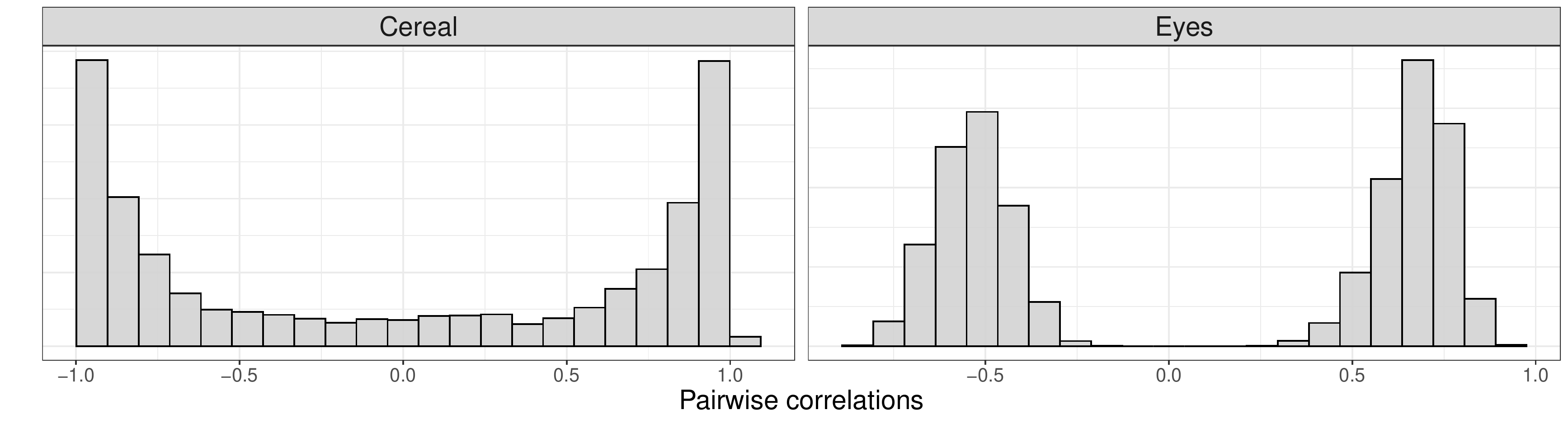}
    \caption{ Histograms of pairwise correlations among covariates across the datasets used in the case study.}
    \label{fig:pairwisecors}
\end{figure}

We evaluate how incorporating covariate dependence structures affects the predictive performance of shrinkage priors using two real high dimensional datasets. The first dataset, Cereal, contains starch content measurements for 15 samples, each with 145 infrared spectral predictors, and is provided by the \texttt{R} package \texttt{chemometrics} \citep{chemometrics}. The second dataset, Eye, includes gene expression levels for 20 genes across 120 samples obtained from microarray experiments on mammalian eye tissue \citep{scheetz_regulation_2006}. It is available in the \texttt{R} package \texttt{flare} \citep{li_flare_2024}. Figure \ref{fig:pairwisecors} displays histograms of the pairwise correlations among covariates. We refrain from displaying the full correlation matrices due to the high dimensionality of the covariate space.

To assess out-of-sample predictive performance, we use ELPD. See Section \ref{sec:eval_metrics} for details. As an independent test set is not available, we rely on exact leave-one-out cross-validation (LOO-CV) \citep{vehtari_practical_2016}. For each fold, the ELPD is computed on the held-out observation and then aggregated across all folds to obtain an overall estimate. We compare the performance of the Horseshoe, Regularized Horseshoe, Dirichlet-Laplace, and R2D2 priors, along with their dependency-aware variants \citep{carvalho_horseshoe_2010, piironen_sparsity_2017, bhattacharya_dirichletlaplace_2015}.

We summarize the results in Table \ref{tab:data_1}, which reports ELPD differences and standard deviations computed via pairwise comparisons against the best-performing model for each dataset. Negative values indicate lower predictive accuracy relative to the best model. 
%
Across real world correlation structures, a consistent pattern emerges: standard shrinkage priors often outperform their dependency-aware counterparts, even in settings with substantial predictor dependence (see Figure \ref{fig:pairwisecors}). While one might expect that explicitly modeling such correlations would enhance predictive performance, the empirical results suggest that standard priors already adapt well to complex structures without requiring explicit dependence modeling.


\begin{table}[t]
\caption{Differences in $\elpd$ and standard deviations for the datasets, computed via pairwise comparisons with the model having the highest $\elpd$ (first column). The initial value is zero, and subsequent columns display negative values that indicate the difference with the best model \citep{looR}. See \cite{vehtari_practical_2016} for details on standard error calculations. Models ending in “O” denote dependency-aware variants and are highlighted in bold blue for emphasis.}
\label{tab:data_1}
\centering
\setlength{\tabcolsep}{2pt}
\begin{threeparttable}
\resizebox{\linewidth}{!}{%
\begin{tabular}{@{}lcccccccc@{}}
\toprule
\textbf{Dataset} &  &  &  &  &  &  &  &  \\
\midrule
Cereal  & \textbf{\textcolor{blue}{DLO}} & DL & D2 & \textbf{\textcolor{blue}{D2O}} & HS & RHS & \textbf{\textcolor{blue}{HSO}} & \textbf{\textcolor{blue}{RHSO}} \\
($n, p ) = (15 , 145$) & 0.0 (0.0) & -4.5 (2.9) & -4.9 (3.4) & -8.7 (5.6) & -17.7 (11.5) & -20.1 (12.5) & -20.7 (11.6) & -27.7 (13.4) \\
\midrule
Eye     & HS & D2 & RHS & \textbf{\textcolor{blue}{HSO}} & \textbf{\textcolor{blue}{RHSO}} & \textbf{\textcolor{blue}{D2O}} & DL & \textbf{\textcolor{blue}{DLO}} \\
($n , p ) = (20, 120$) & 0.0 (0.0) & -2.5 (2.3) & -2.6 (1.1) & -5.8 (1.5) & -10.0 (1.6) & -12.7 (2.3) & -14.2 (4.6) & -22.4 (5.2) \\
\bottomrule
\end{tabular}
}
\end{threeparttable}
\end{table}

\section{Discussion}
\label{sec:discussion}

We propose an extension to the continuous global-local shrinkage priors framework that allows for the inclusion of dependence structures via a correlation matrix $\Omega$. Specifically, we move from the standard independent prior on each coefficient, $b_i \mid \lambda_i, \tau, \sigma \sim \mathcal{N}(0, \lambda_i^2 \tau^2 \sigma^2)$, to a multivariate prior $b \mid \sigma, \tau, \lambda \sim \mathcal{N}\left(0, \sigma^2 \tau^2 \D_\lambda \Omega \D_\lambda\right)$, where $\D_\lambda$ is a diagonal matrix containing the local scales $\lambda_i$, and $\Omega$ is a correlation matrix encoding dependencies among coefficients.

Importantly, our construction imposes no additional constraints on the distributions of the local scales $\lambda_i$, the global scale $\tau$, or their hyperparameters. This preserves the full flexibility of the original shrinkage prior framework. It enables a multivariate generalization of continuous shrinkage priors such as the Horseshoe, without restricting their marginal heavy-tailed behavior or imposing constraints on finite-moment conditions, as was necessary in some earlier approaches \citep{griffin_structured_2024}.

In the absence of knowledge about the true dependence structure, we propose estimating $\Omega$ using information from the design matrix $X$, following ideas akin to Zellner’s $g$ prior. When $p < n$, we use the empirical covariance matrix of $X$. In the high dimensional case ($p > n$), where this estimator becomes unstable, we adopt an automated shrinkage-based estimator that is computationally efficient and enables a fully automated specification of $\Omega$.

Our simulation studies yield mixed results regarding the benefits of incorporating dependence structures. While the intuition that modeling correlations can improve performance is appealing, our results indicate that the benefits are scenario-specific. In particular, improvements in parameter recovery emerge when groups of highly correlated predictors align with the true signal structure. We attribute this to the regularization effect of $\Omega$ on the joint prior for $b$, which tends to mitigate excessive shrinkage of correlated signals. However, when the signal structure does not reflect the correlation pattern, such as in scenarios with randomly distributed nonzero coefficients, this alignment is lost, and the incorporation of $\Omega$ offers no meaningful advantage. Importantly, we find that predictive performance remains stable even when $\Omega$ does not match the signal, suggesting that dependency-aware priors can be safely employed without risk of degradation. If a practitioner is concerned about the limited shrinkage induced by incorporating $\Omega$, hyperparameters of the prior distribution can be adjusted to compensate and the exerted shrinkage can be quantified via the distribution of the effective size of the model. It is worth noting that, for the same set of hyperparameters, dependency-aware variants generally require more computational time to fit.

We emphasize that our approach to incorporating dependency structures via $\Omega$ in the shrinkage prior is only one possible strategy for modeling dependencies. Alternatives include placing a prior directly over $\Omega$ or requiring the user to specify it manually. Placing a prior over $\Omega$ results in a highly overparameterized model, with $p(p-1) / 2$ additional parameters to estimate, increasing data requirements and computational burden. Manual specification shifts a heavy burden onto the user, requiring explicit modeling of $p(p-1) / 2$ correlations, an impractical task in most real world applications. For shrinkage priors to remain practically useful, it is essential that their complexity remains manageable, ideally governed by a small number of hyperparameters, which is satisfied by our approach.

\subsection{Conclusion}

Our study pursued two specific objectives:  First, to develop a streamlined, automated procedure for introducing dependence structures into shrinkage priors. Second, to evaluate whether doing so yields tangible practical benefits. The latter question is often overlooked in the literature, where methodological innovations are proposed without systematically investigating whether they meaningfully improve inference or prediction.

Ultimately, while the incorporation of dependence structures into shrinkage priors is theoretically appealing, our results suggest that in many practical settings, the added complexity is difficult to justify. Standard shrinkage priors, with their simplicity, robustness, and strong predictive performance, often provide a better balance between model flexibility and computational efficiency. We therefore recommend that practitioners prioritize simpler formulations unless there is strong prior evidence of structured correlations among predictors. In short, dependence modeling should be seen not as a default strategy, but as a targeted tool—useful when needed, but costly when applied indiscriminately.

\begin{acks}[Acknowledgments]
Funded by Deutsche Forschungsgemeinschaft (DFG, German Research Foundation) Project 500663361. We acknowledge the computing time provided on the Linux HPC cluster at Technical University Dortmund (LiDO3), partially funded in the course of the Large-Scale Equipment Initiative by DFG Project 271512359. Paul-Christian Bürkner further acknowledges support of the Deutsche Forschungsgemeinschaft (DFG, German Research Foundation) via the Collaborative Research Center 391 (Spatio-Temporal Statistics for the Transition of Energy and Transport) – 520388526. 

The authors would like to thank Yuexi Wang, Luna Fazio, and David Kohns for their thoughtful comments and discussion on earlier versions of the manuscript.
\end{acks}


\section{Appendix}
\label{appendix}

\subsection{Shrinkage priors}

We consider the following shrinkage priors, with default hyperparameter settings detailed below.  

Following the recommendations of \cite{gelman_prior_2006}, we place a Half Student-$t$ prior on the error scale parameter $\sigma$, with $\nu$ degrees of freedom and scale parameter $\eta$. Consistent with the approach of \cite{burkner_brms_2017}, we set $\eta \approx \mathrm{sd}(y)$, as both the prior mean and variance are proportional to $\eta$.

\subsubsection{Beta Prime}

The Beta Prime (BP) prior is specified via:
\[
\begin{aligned} 
b_j \mid \lambda_j, \sigma &\sim \mathcal{N}\left(0,  \lambda_j^2 \sigma^2 \right), \, \,
\lambda_j^2 \sim \mathrm{BetaPrime}( \alpha, \beta), \, \,
\sigma \sim p(\sigma)
\end{aligned}
\]
where $\mathrm{BetaPrime}(\alpha, \beta)$ has density proportional to $\lambda^{\alpha-1}(1+\lambda)^{-\alpha-\beta}$ \citep{bai_large-scale_2019,bai_beta_2021}. 

The suggested default choices are $\alpha \leq 0.5 $ to generate unbounded marginal distributions for $\beta$. We set $\alpha \sim \textrm{Gamma}(1, 2)$ and $\beta = 1$.

\subsubsection{Dirichlet Laplace}

The Dirichlet Laplace (DL) prior induces global-local shrinkage through the following hierarchy:
\[
\begin{aligned}
b_j \mid \psi_j, \phi, \tau, \sigma  &\sim \mathcal{N}\left(0,\,  \psi_j (\phi_j \tau)^2 \sigma^2 \right), \\
\psi_j &\sim \mathrm{Exp}\left(1/2\right), \\
\phi &= (\phi_1, \ldots, \phi_p) \sim \mathrm{Dirichlet}(a_\pi, \ldots, a_\pi), \\
\tau &\sim \mathrm{Gamma}(n a_\pi, 1/2) \\
\sigma  & \sim p(\sigma)
\end{aligned}
\]
where $\mathrm{Exp}(1/2)$ denotes the exponential distribution with rate $1/2$. We set $a_\pi = 0.5 $ as suggested in the Simulation Study by \cite{bhattacharya_dirichletlaplace_2015}.

\subsubsection{Horseshoe}

The Horseshoe prior is defined hierarchically as:
\[
\begin{aligned}
b_j \mid \lambda_j, \tau, \sigma \sim \mathcal{N}\left(0,\, \sigma^2 \lambda_j^2 \tau^2\right), \, \,
\lambda_j \sim \mathcal{C}^+(0,1), \, \,
\tau \sim \mathcal{C}^+(0,1), \, \, \sigma  & \sim p(\sigma)
\end{aligned}
\]
where $\mathcal{C}^+(0,1)$ denotes the half-Cauchy distribution with unit scale \citep{carvalho_horseshoe_2010}. Notice that the Horseshoe prior does not possess hyperparameters others than the one present in the distribution of $\sigma$.

\subsubsection{Normal-Gamma}

The Normal Gamma prior proposed by \cite{griffin_inference_2010} is specified as follows:
\[
\begin{aligned}
\beta_i \mid \psi_i, \sigma &\sim \mathcal{N}(0, \sigma^2 \psi_i^2), \\
\psi_i^2 \mid \lambda, \gamma &\sim \mathrm{Gamma}\left(\lambda, \frac{1}{2\gamma^2}\right), \\
\nu &= 2\lambda\gamma^2, \\
\nu \mid \lambda &\sim \mathrm{InverseGamma}(2, M), \\
\lambda &\sim \mathrm{Exponential}(1).
\end{aligned}
\]

\cite{griffin_inference_2010} argue that centering the prior for $\lambda$ around $1$ introduces variability around the Bayesian LASSO prior. 
In this specification, $\nu$ has expectation $M$, which serves as an informed guess for the squared $\ell_2$-norm of $b$. 
Following this recommendation, we set 
\[
M = \frac{1}{p} \sum_{i=1}^p \hat{b}_i^2,
\]
where $\hat{b}$ is the maximum likelihood estimator (MLE) of $b$ when $X$ has full rank, and the minimum-length least squares estimator when $p > n$ \citep{brown_measurement_1994}. Setting $ \lambda = 1/2 $, $M = 1$ and $1 / \nu \sim \mathrm{Gamma}(1/2, 1)$ results in the Horseshoe prior \citep{griffin_bayesian_2021}.

\subsubsection{R2D2}

The R2D2 prior places a distribution on the proportion of explained variance $R^2$, and allocates the total prior variance $\omega^2$ among coefficients using a Dirichlet distribution \cite{zhang_bayesian_2020}. 
Specifically, a Beta prior is assigned to $R^2$, which induces a Beta Prime prior on $\omega^2$ with parameters $a_1, a_2 > 0$. 

The total variance $\omega^2$ is decomposed across coefficients using proportions $\phi = (\phi_1, \dots, \phi_p)$, where $\phi \sim \mathrm{Dirichlet}(\alpha)$ and $\lambda_i^2 = \phi_i \omega^2$. Originally, \cite{zhang_bayesian_2020} specified a Double Exponential prior for the coefficients, while \cite{aguilar_intuitive_2023} proposed a Normal prior. Following the recommendations of \cite{aguilar_intuitive_2023}, we adopt a Normal prior specification and set the hyperparameters $\alpha=(a_\pi,...,a_\pi)$ with $a_\pi = 0.25$, $a_2 = 0.5$ to ensure unbounded marginal distributions and heavy-tailed behavior. \cite{aguilar_generalized_2025} show that other decompositions of variance are available, however we have only considered the Dirichlet in our experiments. 

The full hierarchical structure of the R2D2 prior is:
\[
\begin{aligned}
b_j \mid \lambda_j^2, \sigma^2 &\sim \mathcal{N}\left(0,\, \sigma^2 \lambda_j^2\right), \\
\lambda_j^2 &= \phi_j \omega^2, \\
\phi &\sim \mathrm{Dirichlet}(a_\pi, \dots, a_\pi), \\
\omega^2 &\sim \mathrm{BetaPrime}(a_1, a_2) \, \sigma  \sim p(\sigma).
\end{aligned}
\]

\subsection{Bounds}

In this subsection we prove the bounds presented in Section \ref{subsec:cond_means} shown in \eqref{eq:scaled-norm-bound}. Our objective is to bound the spectral norm of $Q_\Omega^{-1} - Q_I^{-1}$, where

\begin{equation}
\begin{aligned}
Q_\Omega &= X'X + D_\lambda^{-1} \Omega^{-1} D_\lambda^{-1}, \\
Q_I &= X' X + D_\lambda^{-2}.      
\end{aligned}
\end{equation}

In the following let $\lambda_1 = \max_i \lambda_i,  \lambda_p = \min_i \lambda_i, \nu_1 = \lambda_{\max}(X' X), \nu_p = \lambda_{\min}(X' X), \omega_1 = \lambda_{\max}(\Omega), \omega_p = \lambda_{\min}(\Omega)$ where $\lambda_{\max}(A)$ and $\lambda_{\min}(A)$ denote the maximum and minimum eigenvalues of $A$ respectively. 

We make use of the following:

\begin{itemize}
    \item Let $A \in \mathbb{R}^{p \times p}$ be a real symmetric matrix and $x \in \mathbb{R}^p$ a nonzero vector. Then its spectral norm is given by
    \begin{equation}
    \|A\|_2 = \max_{x \neq 0} \frac{\|Ax\|_2}{\|x\|_2} = \max_{i} |\lambda_i(A)|,
    \end{equation}
    where $\lambda_i(A)$ denotes the $i$-th eigenvalue of $A$. In particular, if $A$ is positive semidefinite, the spectral norm equals its largest eigenvalue.
    
    \item The  resolvent identity, which states that for invertible matrices $A$ and $B$:

    \begin{equation}
        A^{-1} - B^{-1} = A^{-1}(B - A)B^{-1}.
    \end{equation}
    
    \item The submultiplicative property of the spectral norm:
    \begin{equation}
        \|AB\|_2 \leq \|A\|_2  \|B\|_2,
    \end{equation}
    
    \item If $A$ and $C$ are invertible then:
    
    \begin{equation}
    \label{eq:reverse_inequality}
    \|ABC\|_2 \geq \|A^{-1}\|_2^{-1} \|B\|_2 \cdot \|C^{-1}\|_2^{-1}.
    \end{equation}

    \item Rayleigh characterization. Let $A \in \mathbb{R}^{p \times p}$ be a real symmetric matrix. Then its smallest and largest eigenvalues can be characterized as \citep{anderson_introduction_2003}
    \begin{equation}
    \label{eq:rayleigh_char}
    \lambda_{\min}(A) = \min_{\|x\|_2 = 1} x' A x, \qquad
    \lambda_{\max}(A) = \max_{\|x\|_2 = 1} x' A x.
    \end{equation}
       
    \item  Weyl’s inequality for symmetric matrices $A$ and $B$ \citep{weyl_asymptotische_1912}:
    \begin{equation}
    \begin{aligned}
    \lambda_{\min}(A + B) &\geq \lambda_{\min}(A) + \lambda_{\min}(B), \\
    \lambda_{\max}(A + B) &\leq \lambda_{\max}(A) + \lambda_{\max}(B).  
    \end{aligned}
    \end{equation}
 
\end{itemize}

We begin by proving the upper bound. An application of the resolvent identity yields

\begin{equation}
\label{eq:bound_resolvent}
Q_\Omega^{-1} - Q_I^{-1}
= Q_\Omega^{-1}(Q_I - Q_\Omega) Q_I^{-1}
= Q_\Omega^{-1} D_\lambda^{-1}(\Omega^{-1} - I) D_\lambda^{-1} Q_I^{-1}.
\end{equation}

Taking spectral norms and applying the submultiplicative property:
\[
\| Q_\Omega^{-1} - Q_I^{-1} \|_2
\leq \| Q_\Omega^{-1} \|_2 \cdot \| D_\lambda^{-1} (\Omega^{-1} - I) D_\lambda^{-1} \|_2 \cdot \| Q_I^{-1} \|_2.
\]

We now proceed to bound each term. First, applying the submultiplicative property to the norm in the middle of the right hand, we have 

\begin{equation}
\label{eq:upper_bound_middle}
\| D_\lambda^{-1} (\Omega^{-1} - I) D_\lambda^{-1} \|_2
\leq \| D_\lambda^{-1} \|_2^2  \| \Omega^{-1} - I \|_2
= \frac{1}{\lambda_p^2}  \| \Omega^{-1} - I \|_2.
\end{equation}

Next, we bound the spectral norms of $Q_\Omega^{-1}$ and $Q_I^{-1}$. Since $Q_\Omega$ and $Q_I$ are symmetric positive definite, the spectral norms of $Q_{\Omega}^{-1}$ and $Q_I^{-1}$ are given by the reciprocals of their smallest eigenvalues of $Q_\Omega$ and $Q_I$ respectively. By Weyl’s inequality:
\begin{equation}
\begin{aligned}
    \lambda_{\min}(Q_\Omega) = \lambda_{\min}\left( X'X + D_\lambda^{-1} \Omega^{-1} D_\lambda^{-1}  \right)
&\geq \lambda_{\min}(X'X) + \lambda_{\min}(D_\lambda^{-1} \Omega^{-1} D_\lambda^{-1}) \\
&\geq  \nu_p + \frac{1}{\lambda_1^2 \omega_1}.
\end{aligned}
\end{equation}

In the last line we have used 
\begin{equation*}
    \lambda_{\min}(D_\lambda^{-1} \Omega^{-1} D_\lambda^{-1}) 
    = \frac{1}{\lambda_{\max}(D_\lambda \Omega  D_\lambda)} \geq \frac{1}{\lambda_{\max}(D_\lambda^2) \lambda_{\max}(\Omega) } = \frac{1}{\lambda_1^2  \omega_1},
\end{equation*}
which follows from the submultiplicative property of the spectral norm.

Using the same argument we obtain
\begin{equation*}
\lambda_{\min}(Q_I) \geq \nu_p + \frac{1}{\lambda_1^2}.
\end{equation*}

Therefore 

\begin{equation}
\label{eq:upper_bound_left_right}
    \begin{aligned}
        \| Q_\Omega^{-1} \|_2 \leq \left( \nu_p + \frac{1}{\lambda_1^2 \omega_1} \right)^{-1}, \ \ 
\| Q_I^{-1} \|_2 \leq \left( \nu_p + \frac{1}{\lambda_1^2} \right)^{-1}.
    \end{aligned}
\end{equation}

Combining \eqref{eq:upper_bound_middle} and  \eqref{eq:upper_bound_left_right} yields

\begin{equation}
\begin{aligned}
\| Q_\Omega^{-1} - Q_I^{-1} \|_2 
\leq
\frac{\| \Omega^{-1} - I \|_2}{\lambda_p^2 \left( \nu_p + \frac{1}{\lambda_1^2 \omega_1} \right) \left( \nu_p + \frac{1}{\lambda_1^2} \right)}
\end{aligned}
\end{equation}

We proceed to prove the lower bound. From the identity:
\[
Q_\Omega^{-1} - Q_I^{-1} = Q_\Omega^{-1} D_\lambda^{-1}(\Omega^{-1} - I) D_\lambda^{-1} Q_I^{-1},
\]
we apply the reverse inequality for products of invertible matrices shown in \ref{eq:reverse_inequality} to get 

\begin{equation}
\label{eq:lower_reverse_ineq}
\| Q_\Omega^{-1} - Q_I^{-1} \|_2
\geq \| Q_\Omega \|_2^{-1}  \| D_\lambda^{-1}(\Omega^{-1} - I) D_\lambda^{-1} \|_2  \| Q_I \|_2^{-1}. 
\end{equation}

To lower bound $ \| Q_\Omega \|_2^{-1} $ and $\| Q_I \|_2^{-1}$ we make use of Weyl's inequality on $\| Q_\Omega \|_2 $ and $\| Q_I \|_2$:

\begin{equation}
\begin{aligned}
    \lambda_{\max}(Q_\Omega) = \lambda_{\max}\left( X'X + D_\lambda^{-1} \Omega^{-1} D_\lambda^{-1}  \right)
&\leq \lambda_{\max}(X'X) + \lambda_{\max}(D_\lambda^{-1} \Omega^{-1} D_\lambda^{-1}) \\
&\leq  \nu_1 + \frac{1}{\lambda_p^2 \omega_p}.
\end{aligned}
\end{equation}

In the last line we have used that 

\begin{equation*}
\begin{aligned}
    \lambda_{\max}(D_\lambda^{-1} \Omega^{-1} D_\lambda^{-1}) = \frac{1}{ \lambda_{\min}(D_\lambda \Omega D_\lambda)  } \leq  \frac{1}{ \lambda_{\min}(D_\lambda^2)  \lambda_{\min}(\Omega)} = \frac{1}{ \lambda_p^2 \omega_p}.
\end{aligned}
\end{equation*}

This statement follows from the Rayleigh characterization shown in \ref{eq:rayleigh_char}. Given that $\lambda_{\min}(D_\lambda \Omega D_\lambda) = \min_{\|x\|_2 = 1} x' D_\lambda \Omega D_\lambda x$ and letting $y = D_\lambda x$ leads to $ \| y \|^2 = y'y = x' D_\lambda^2 x$. This implies that $ \lambda_{\min}(D_\lambda^2) \leq \| y \|^2 \leq \lambda_{\max}(D_\lambda^2)$. Therefore

\begin{equation*}
    \begin{aligned}
        x' D_\lambda \Omega D_\lambda x \geq  y' \Omega y \geq \lambda_{\min}(\Omega) y'y.
    \end{aligned}
\end{equation*}
Taking the minimum yields 
\begin{equation*}
    \begin{aligned}
        \lambda_{\min} (D_\lambda \Omega D_\lambda ) \geq \lambda_{\min}(\Omega) \lambda_{\min}(D_\lambda^2) = \omega_p \lambda_p^2.
    \end{aligned}
\end{equation*}

A similar arguments produces 

\begin{equation}
\begin{aligned}
    \lambda_{\max}(Q_I) \leq  \nu_1 + \frac{1}{\lambda_p^2 }.
\end{aligned}
\end{equation}

Therefore

\begin{equation}
\label{eq:lower_bound_left_right}
\| Q_\Omega \|_2 \leq \nu_1 + \frac{1}{\lambda_p^2 \omega_p}, \qquad
\| Q_I \|_2 \leq \nu_1 + \frac{1}{\lambda_p^2},
\end{equation}

We proceed to bound $ \| D_\lambda^{-1}(\Omega^{-1} - I) D_\lambda^{-1} \|_2$ in line \ref{eq:lower_reverse_ineq}.
By applying the reverse inequality for products of invertible matrices we obtain
\begin{equation}
\begin{aligned}
\label{eq:lower_bound_middle}
\| D_\lambda^{-1}(\Omega^{-1} - I) D_\lambda^{-1} \|_2
\geq \frac{1}{\lambda_1^2} \cdot \| \Omega^{-1} - I \|_2,    
\end{aligned}    
\end{equation}

Combining \eqref{eq:lower_bound_left_right} and \eqref{eq:lower_bound_middle} results in

\[
\| Q_\Omega^{-1} - Q_I^{-1} \|_2 \| 
\geq
\frac{ \| \Omega^{-1} - I \|_2}{\lambda_1^2 \left( \nu_1 + \frac{1}{\lambda_p^2 \omega_p} \right)\left( \nu_1 + \frac{1}{\lambda_p^2} \right)}
\]

\subsection{Correlation Matrices}

We have used the following correlation matrix structures in the main text. For further details on these and related constructions, see \citet{rue_gaussian_2005} and \citet{box2015time}.

\subsubsection{Autoregressive of Order 1}

The autoregressive correlation matrix of order 1, AR(1), assumes that the correlation between two observations decays exponentially with their distance. Specifically, for observations $i$ and $j$, the $(i,j)$-th entry of the matrix is given by
\begin{equation}
\text{Corr}(i,j) = \rho^{|i-j|},
\end{equation}
where $\rho \in (-1, 1)$ is the autocorrelation parameter.

\subsubsection{Moving Average of order 1}
The moving average correlation matrix of order 1 MA(1) captures correlations between immediate neighbors. The $(i,j)$-th entry of the matrix is
\[
\text{Corr}(i,j) =
\begin{cases}
1 & \text{if } i = j, \\
\rho & \text{if } |i-j| = 1, \\
0 & \text{otherwise},
\end{cases}
\]
where $\rho \in (-1,1)$ measures the strength of correlation between adjacent observations.

\subsubsection{Moving Average of order 2}
The moving average correlation matrix of order 2 MA(2) extends the MA(1) structure to also include second neighbors. The $(i,j)$-th entry of the matrix is
\[
\text{Corr}(i,j) =
\begin{cases}
1 & \text{if } i = j, \\
\rho_1 & \text{if } |i-j| = 1, \\
\rho_2 & \text{if } |i-j| = 2, \\
0 & \text{otherwise},
\end{cases}
\]
where $\rho_1, \rho_2 \in (-1,1)$ and control the correlation between first and second neighbors, respectively. Additionally it is required that $\rho_1 \pm \rho_2 <1$. To parametrize the MA(2) in terms of only $\rho \in (-1,1)$ we have set $\rho_1 = \rho$ and $\rho_2 = (1-\rho)\rho$.

\subsection{Ledoit-Wolf estimator}

\cite{ledoit_well-conditioned_2004} propose the following linear shrinkage estimator
\begin{equation}
\Sigma^* =  \varphi_1 I + \varphi_2 S = \frac{\beta^2}{ \delta^2} \mu I + \frac{\alpha^2}{\delta^2} {S}.
\end{equation}
where $\delta^2 = \alpha^2 + \beta^2$ and $\Sigma^*$ minimizes the expected quadratic loss $\mathbb{E} \left\| \tilde{\Sigma} - \Sigma_X \right\|^2_F$ subject to $\tilde{\Sigma}= \varphi_1 I + \varphi_2 S$, with respect to nonrandom coefficients $\varphi_1, \varphi_2$. Here, $\|\cdot \|_F$ denotes the scaled Frobenius norm, i.e., $\| A \|_F = \text{tr}(A' A)/p$. The shrinkage parameters in Equation are computed as follows:
\begin{equation}
\mu = \frac{\text{tr}(S)}{p}, \quad \alpha^2 = \| \Sigma_X - \mu I \|_F^2, \quad \beta^2 = \mathbb{E} \| S - \Sigma_X \|^2_F.
\end{equation}

Since $\Sigma_X$ is unknown, consistent estimators for $\mu$, $\alpha^2$, and $\beta^2$ must be used to construct a consistent estimator $S^*$ of $\Sigma^*$. 

\begin{proposition}
Let $ X \in \mathbb{R}^{n \times p} $ be a design matrix, and define
\begin{equation*}
\begin{aligned}
m_n &= \frac{\mathrm{tr}(S)}{p}, \quad 
d_n^2 = \| S - m_n I \|_F^2, \\
b_n^2 &= \min \left( \frac{1}{n^2} \sum_{k=1}^n \left\| x_{k\cdot} x_{k\cdot}^\top - S \right\|_F^2,\ d_n^2 \right), \quad 
a_n^2 = d_n^2 - b_n^2,
\end{aligned}
\end{equation*}
where $ x_{k\cdot} $ denotes the $ k $-th row of $ X $, and $ S $ is the sample covariance matrix.

Then:
\begin{enumerate}
    \item $ m_n $, $ d_n^2 $, and $ b_n^2 $ are consistent (in quadratic mean) estimators of $ \mu $, $ \delta^2 $, and $ \beta^2 $, respectively.
    
    \item The resulting linear shrinkage estimator
    \begin{equation}
    S^* = \frac{b_n^2}{d_n^2} m_n I + \frac{a_n^2}{d_n^2} S
    \end{equation}
    is a consistent estimator of the Ledoit–Wolf shrinkage estimator $ \Sigma^* $.
\end{enumerate}
\end{proposition}
See \citet{ledoit_well-conditioned_2004} for proof and further details.

\subsection{Simulations: Further results}

\subsubsection{Coverage}

Tables \ref{tab:bma1_fixed3} and \ref{tab:bar1_fixed3} summarize the coverage performance of various shrinkage priors under two correlation levels ($\rho = 0.5$ and $\rho = 0.95$), with fixed $p = 250$ and high signal strength ($R^2 = 0.8$) when using 95\% posterior credibility intervals using the BMA1 and BAR1 structures respectively. Models with an “O” suffix correspond to dependency-aware priors. The high-$R^2$ setting ensures that signal is present and identifiable, making this a meaningful context to evaluate how different priors handle multicollinearity. When $R^2$ is low, most of the variation in the response remains unexplained, making it difficult to discern meaningful signals from noise regardless of the prior. 

Regarding the results of structure BMA1 shown in Table \ref{tab:bma1_fixed3}, all models maintain high coverage and specificity across settings. The most striking differences emerge in sensitivity and nonzero coverage, particularly as $\rho$ increases. While base priors suffer sharp declines in power at high $\rho$ (for instance the HS drops from 0.729 to 0.352 while HSO moves from 0.921 to 0.906), dependency-aware priors remain robust, consistently recovering signals with high probability. This echoes the results of parameter recovery that we show in Section \ref{sec:methods}.

In contrast to the BMA1 structure, where dependency-aware priors delivered clear improvements under high collinearity, the BAR1 Fixed design proves more challenging. At $\rho = 0.5$, all models achieve high coverage and specificity, but sensitivity remains modest—generally between 0.2 and 0.5. Notably, the dependency-aware variants often underperform their unstructured counterparts, suggesting that incorporating the BMA1-based precision matrix can over-shrink when the signal is not well aligned with the imposed structure. As $\rho$ increases to 0.95, sensitivity deteriorates across all models, falling below 0.15. Even dependency-aware priors, which excelled under BMA1, fail to recover meaningful signal here.

\begin{figure}[t!]
    \centering
    \includegraphics[width=1\linewidth]{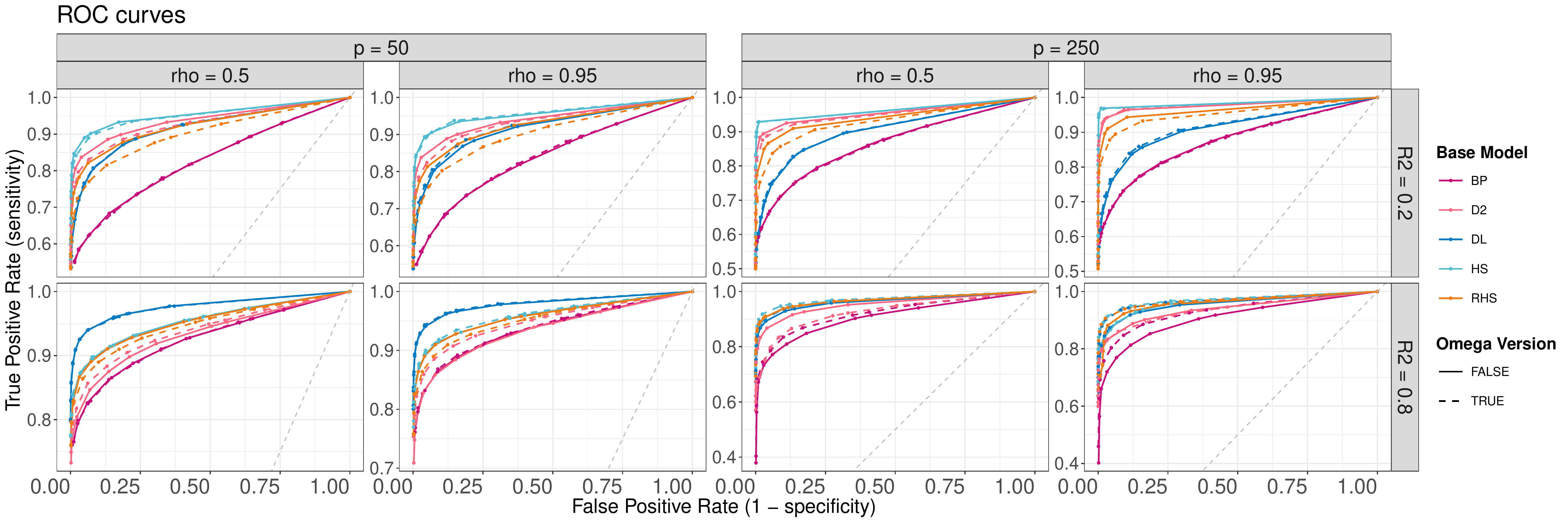}
    \vspace{-0.5cm}  
    \caption{ \textbf{ROC curves for the BMA1 with fixed block signals ($p \in \{50, 250\}$, $R^2 \in \{0.2, 0.8\}$).} We compare standard shrinkage priors (solid lines) and their dependency-aware counterparts (dashed lines). Models with an "O" suffix incorporate the BMA1 structure to induce dependency-aware shrinkage. Performance gains are most visible at high $\rho$ and high $R^2$, where structured shrinkage improves signal recovery.}
    \label{fig:roc_blockedma1_fixed}
\end{figure}

\begin{figure}[t!]
    \centering
    \includegraphics[width=1\linewidth]{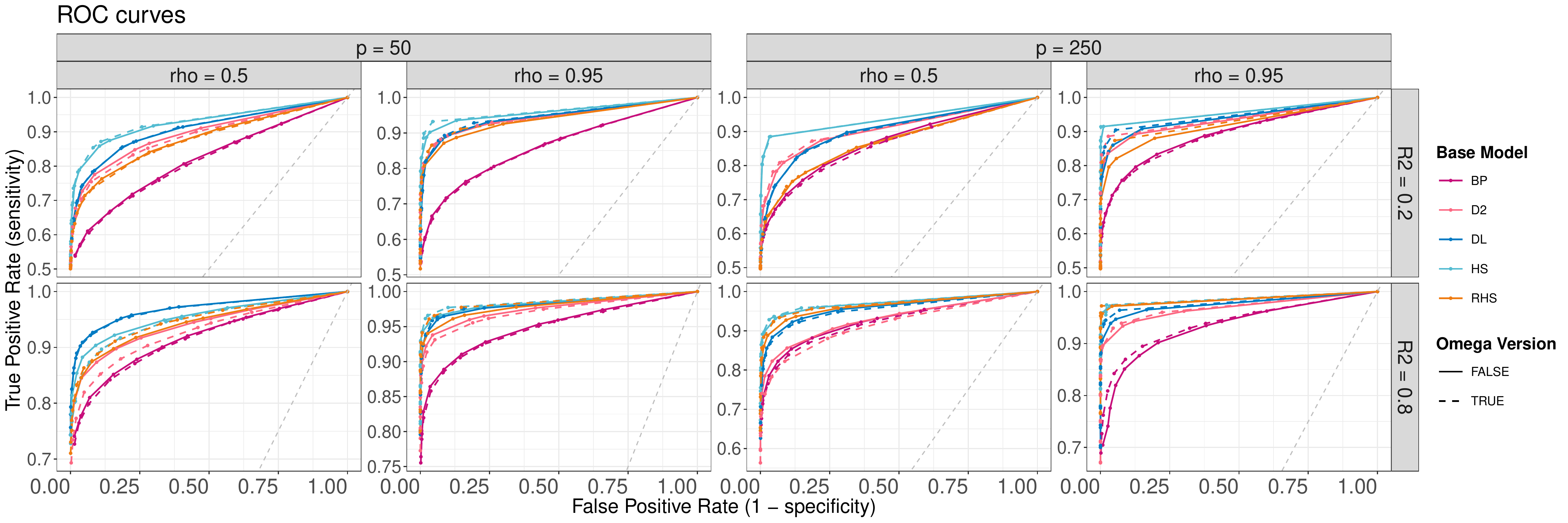}
    \vspace{-0.5cm}  
    \caption{ \textbf{ROC curves for the BAR1 with fixed block signals ($p \in \{50, 250\}$, $R^2 \in \{0.2, 0.8\}$).} We compare standard shrinkage priors (solid lines) and their dependency-aware counterparts (dashed lines). Models with an "O" suffix incorporate the BAR1 structure to induce dependency-aware shrinkage. As a result, performance is comparable or slightly degraded relative to the standard versions.}    
    \label{fig:roc_blockedar1_fixed}
\end{figure}

Figures~\ref{fig:roc_blockedma1_fixed} and~\ref{fig:roc_blockedar1_fixed} show ROC curves under the BMA1 and BAR1 structures, respectively, across varying levels of correlation ($\rho \in {0.5, 0.95}$), dimensionality ($p \in {50, 250}$), and signal strength ($R^2 \in {0.2, 0.8}$). As before, each standard shrinkage prior is plotted alongside its dependency-aware version (dashed line of the same color).

Under the BMA1 structure, dependency-aware priors consistently outperform or match their unstructured counterparts, especially at high $\rho$ and high $R^2$. In these settings, dependency-aware shrinkage priors improve true positive rates without sacrificing specificity, resulting in ROC curves that dominate or closely track above their base versions. In contrast, under the BAR1 structure, the dependency-aware variants perform similarly or slightly worse than the standard priors. The ROC curves largely overlap, and in some cases  the dependency-aware versions lag behind. This reinforces the earlier observation that the utility of structured priors is highly design dependent: they are most effective when the imposed dependence structure is aligned with the true generative mechanism (as in BMA1), but offer limited or no benefit when the structure is mismatched (as in BAR1).

\begin{table}[!h]
\centering
\caption{\textbf{Coverage metrics under the BMA1 structure with fixed block signals ($p = 250$, $R^2 = 0.8$).} We compare shrinkage priors across two correlation levels ($\rho = 0.5$ and $\rho = 0.95$). Models with an "O" suffix incorporate the BMA1 structure to induce dependency-aware shrinkage.}
\label{tab:bma1_fixed3}
\resizebox{\linewidth}{!}{
\begin{tabular}{ccccccc}
\toprule
Model ID & Coverage & Specificity & Sensitivity (Power) & Avg. CI Width & Coverage Zero & Coverage Nonzero \\
\midrule
\multicolumn{7}{l}{\textbf{Condition: $\rho = 0.5$}} \\
BP   & 0.984 & 0.996 & 0.831 & 1.032 & 0.996 & 0.696 \\
BPO  & 0.989 & 0.997 & 0.990 & 1.028 & 0.997 & 0.794 \\
D2   & 0.961 & 1.000 & 0.067 & 0.972 & 1.000 & 0.029 \\
D2O  & 0.971 & 0.998 & 0.733 & 1.156 & 0.998 & 0.321 \\
DL   & 0.989 & 0.999 & 0.860 & 0.869 & 0.999 & 0.733 \\
DLO  & 0.992 & 0.999 & 0.988 & 0.819 & 0.999 & 0.817 \\
HS   & 0.987 & 1.000 & 0.729 & 0.603 & 1.000 & 0.688 \\
HSO  & 0.994 & 1.000 & 0.921 & 0.669 & 1.000 & 0.850 \\
RHS  & 0.990 & 1.000 & 0.721 & 0.694 & 1.000 & 0.756 \\
RHSO & 0.993 & 1.000 & 0.942 & 0.758 & 1.000 & 0.825 \\
\midrule
\multicolumn{7}{l}{\textbf{Condition: $\rho = 0.95$}} \\
BP   & 0.974 & 0.995 & 0.569 & 0.916 & 0.995 & 0.463 \\
BPO  & 0.984 & 0.997 & 0.983 & 0.891 & 0.997 & 0.656 \\
D2   & 0.960 & 1.000 & 0.065 & 0.740 & 1.000 & 0.000 \\
D2O  & 0.967 & 1.000 & 0.865 & 1.008 & 1.000 & 0.181 \\
DL   & 0.976 & 0.999 & 0.494 & 0.787 & 0.999 & 0.419 \\
DLO  & 0.989 & 1.000 & 0.967 & 0.674 & 1.000 & 0.735 \\
HS   & 0.975 & 1.000 & 0.352 & 0.457 & 1.000 & 0.371 \\
HSO  & 0.991 & 1.000 & 0.906 & 0.550 & 1.000 & 0.781 \\
RHS  & 0.967 & 0.993 & 0.310 & 0.557 & 0.993 & 0.346 \\
RHSO & 0.991 & 1.000 & 0.940 & 0.636 & 1.000 & 0.769 \\
\bottomrule
\end{tabular}}
\end{table}

\begin{table}[!h]
\centering
\caption{\textbf{Coverage metrics under the BAR1 structure with fixed block signals ($p = 250$, $R^2 = 0.8$).} We compare shrinkage priors across two correlation levels ($\rho = 0.5$ and $\rho = 0.95$). Models with an "O" suffix incorporate the BMA1 structure to induce dependency-aware shrinkage.}
\label{tab:bar1_fixed3}
\resizebox{\linewidth}{!}{
\begin{tabular}{ccccccc}
\toprule
Model ID & Coverage & Specificity & Sensitivity (Power) & Avg. CI Width & Coverage Zero & Coverage Nonzero \\
\midrule
\multicolumn{7}{l}{\textbf{Condition: $\rho = 0.5$}} \\
BP   & 0.987 & 0.995 & 0.506 & 2.358 & 0.995 & 0.802 \\
BPO  & 0.986 & 0.996 & 0.281 & 2.645 & 0.996 & 0.746 \\
D2   & 0.996 & 1.000 & 0.492 & 2.385 & 1.000 & 0.915 \\
D2O  & 0.993 & 1.000 & 0.158 & 2.869 & 1.000 & 0.817 \\
DL   & 0.993 & 0.999 & 0.423 & 1.965 & 0.999 & 0.863 \\
DLO  & 0.990 & 0.999 & 0.275 & 2.190 & 0.999 & 0.769 \\
HS   & 0.993 & 1.000 & 0.323 & 1.469 & 1.000 & 0.823 \\
HSO  & 0.989 & 1.000 & 0.233 & 1.505 & 1.000 & 0.735 \\
RHS  & 0.994 & 1.000 & 0.377 & 1.720 & 1.000 & 0.869 \\
RHSO & 0.989 & 1.000 & 0.212 & 1.688 & 1.000 & 0.740 \\
\midrule
\multicolumn{7}{l}{\textbf{Condition: $\rho = 0.95$}} \\
BP   & 0.969 & 0.973 & 0.142 & 6.425 & 0.973 & 0.858 \\
BPO  & 0.987 & 0.995 & 0.092 & 7.523 & 0.995 & 0.788 \\
D2   & 1.000 & 1.000 & 0.031 & 3.554 & 1.000 & 0.994 \\
D2O  & 0.996 & 1.000 & 0.062 & 5.119 & 1.000 & 0.894 \\
DL   & 0.998 & 1.000 & 0.040 & 3.422 & 1.000 & 0.948 \\
DLO  & 0.994 & 1.000 & 0.042 & 3.935 & 1.000 & 0.863 \\
HS   & 0.985 & 1.000 & 0.065 & 1.363 & 1.000 & 0.629 \\
HSO  & 0.984 & 1.000 & 0.056 & 1.493 & 1.000 & 0.606 \\
RHS  & 0.992 & 1.000 & 0.042 & 1.571 & 1.000 & 0.796 \\
RHSO & 0.989 & 1.000 & 0.050 & 1.744 & 1.000 & 0.725 \\
\bottomrule
\end{tabular}}
\end{table}


\bibliographystyle{ba}
\bibliography{bib/refs}

\end{document}